\documentclass[preprintnumbers,nofootinbib,
superscriptaddress,amsmath,floatfix,prd]{revtex4}

\usepackage{amssymb}  % \gtrsim, \geqslant, etc: see amsguide.ps
\usepackage{graphicx,subfigure}
\usepackage{exscale}
\usepackage{textcomp}
\usepackage{enumerate}
\usepackage{color}

\usepackage[normalem]{ulem}  % \sout{old text} for strikeout

\newcommand{\non}{\nonumber\\}

\newcommand{\be}{\begin{equation}}
\newcommand{\ee}{\end{equation}}
\newcommand{\bea}{\begin{eqnarray}}
\newcommand{\eea}{\end{eqnarray}}
\newcommand{\ba}[1]{\begin{array}{#1}}
\newcommand{\ea}{\end{array}}

\newcommand{\uk}{\hat{\mathbf{k}}}

\begin{document}

\title{Superfluid two-stream instability in a microscopic model}

\author{Andreas Schmitt}
\email{aschmitt@hep.itp.tuwien.ac.at}
\affiliation{Institut f\"{u}r Theoretische Physik, Technische Universit\"{a}t Wien, 1040 Vienna, Austria}

\date{3 March 2014}

\begin{abstract}

The superflow in a superfluid is bounded from above by Landau's critical velocity. Within a microscopic bosonic model, I show that below this critical velocity there is 
a dynamical instability that manifests itself in an imaginary sound velocity and that is reminiscent of the two-stream instability in electromagnetic
plasmas. I compute the onset of this instability and its full angular structure in a relativistic, uniform superfluid for all temperatures. 
At weak coupling, the instability only operates in a very small region in the phase diagram of temperature and superflow. Varying the 
coupling of the model suggests that the effect is more prominent at strong coupling and thus could be important for superfluids in compact stars 
and in the laboratory.

\end{abstract}

%\pacs{12.38.Mh,24.85.+p}

\maketitle

%\tableofcontents

\section{Introduction}
\label{intro}

Many hydrodynamic properties of superfluids can be understood in terms of the two-fluid 
picture \cite{tisza38,landau41,1982PhLA...91...70K,1982ZhETF..83.1601L,1992PhRvD..45.4536C}. 
In this picture, the superfluid consists of a single fluid at zero 
temperature, and is a mixture of a superfluid and a normal-fluid component at nonzero temperatures below the critical temperature. 
One of the consequences is the existence of a second sound mode. 

Landau has argued that a superfluid dissipates energy even at zero temperature if the velocity of the superfluid, for instance with respect to a capillary,
is sufficiently large. The resulting critical velocity manifests itself in the onset of negative quasiparticle energies. At nonzero temperatures, 
this argument is valid even in an infinite system, i.e., without any interaction with the walls of the capillary: 
now there is a second relevant rest frame, the one of the normal fluid. If the relative velocity between the two fluid components becomes 
sufficiently large, the quasiparticle energies become negative and one thus finds a temperature-dependent critical velocity. 
%In the model that 
%I will use here, this critical velocity was computed recently in Ref.\ \cite{Alford:2013koa}.

In this paper, I discuss an additional critical velocity that manifests itself in complex sound velocities, indicating an exponentially 
growing amplitude of one of the sound waves. The sound modes 
are computed in the dissipationless limit and with the help of the linearized hydrodynamic equations. Therefore, the calculation will only indicate the onset of 
the instability, not its  temporal evolution or any damping, for which non-linear effects and dissipation would have to be taken into account. 
One of the main results will be the identification of the unstable regions in the phase diagram.   
The microscopic model is given by a field theory for a complex scalar field $\varphi$ with a $\varphi^4$ interaction,
whose parameters are the boson mass and the coupling constant. The calculations are performed within the self-consistent two-particle irreducible formalism (2PI)
\cite{Luttinger:1960ua,Baym:1962sx,Cornwall:1974vz} in the Hartree approximation, following Ref.\ \cite{Alford:2013koa}. Besides the parameters
in the Lagrangian, the system will be characterized by chemical potential, temperature, and the relative flow between superfluid and normal fluid. 
This flow is assumed to be uniform. 

It turns out that the instability is analogous to the so-called two-stream instability that plays an important role in electromagnetic plasmas and which, 
in that context, is sometimes called Farley-Buneman instability \cite{Buneman:1959zz,1963PhRvL..10..279F,2001AmJPh..69.1262A,2010PhPl...17l0501B}. 
The relevance of the two-stream instability for superfluids has been pointed out in Ref.\ \cite{2004MNRAS.354..101A}. 
Of particular interest for the present work is the relativistic discussion 
of the two-stream instability in Ref.\ \cite{Samuelsson:2009up}. In this reference, the instability is discussed in a general two-fluid formalism, 
making no reference to superfluidity or to any microscopic model. I will show that in the present field-theoretical calculation some of the 
qualitative features of these general results are reproduced. 
%I go beyond that work by starting from a microscopic theory and discussing the full
%angular dependence of the instability. 

Relativistic superfluids are likely to be present in compact stars. In this extreme high-density environment, nuclear matter can become a superfluid
via Cooper pair condensation of neutrons. Moreover, if quark matter is present in the core of the star, it may be superfluid in the color-flavor-locked
phase \cite{Alford:1998mk,Alford:2007xm}, where quarks of all three colors and flavors form Cooper pairs and which spontaneously breaks the global 
$U(1)$ associated with baryon number 
conservation. Consequently, the superfluid two-stream instability discussed here may be important for the physics of compact stars, for instance
in the context of pulsar glitches \cite{2004MNRAS.354..101A}. The present work is not restricted to high-energy superfluids, since the model can 
continuously extrapolate between ultra-relativistic and non-relativistic limits by varying the boson mass. Therefore, the results may also be interesting 
for superfluids in cold atomic gases, where second sound \cite{2013arXiv1302.2871S} and critical velocities \cite{1999PhRvL..83.2502R} have been measured, or
for superfluid helium, where experiments have shown that dissipationless superfluidity is lost for superfluid velocities far below Landau's
critical velocity, for instance through vortex nucleation \cite{1985PhRvL..55.2704A,1992PhRvL..68.2624A}. 

%The two-stream instability is closely related to the interaction between the two fluids. In the context of a superfluid, this interaction 
%can be characterized by the entrainment between the two fluids corresponding to the entropy current and the charge current. (If dissipation is neglected,
%both of these currents are conserved.) More entrainment appears to lead to a more severe instability \cite{Samuelsson:2009up}. In the present 
%field-theoretical model, the entrainment coefficient has been computed in Ref.\ \cite{2013PhRvD..87f5001A,Alford:2013koa}, and thus the results of the 
%present paper can be related to the previous results for entrainment. 

The paper is organized as follows.  In Sec.\ \ref{sec:formalism}, I will briefly explain the calculation of the sound velocities and the microscopic
model. The two-stream instability is analyzed in detail in Sec.\ \ref{sec:twostream}, with various aspects being discussed in the 
subsections of this part: the absence of the instability at zero temperature, Sec.\ \ref{sec:zeroT}; the typical manifestation of the two-stream instability
in the upstream direction for an intermediate temperature, Sec.\ \ref{sec:nonzeroT}; the full angular dependence of the instability, Sec.\ \ref{sec:angles}; 
the dependence on the boson mass and the coupling strength, Sec.\ \ref{sec:dependence}; and the phase diagram in the plane of temperature and superfluid velocity, 
Sec.\ \ref{sec:phasediagram}. I give some conclusions in Sec.\ \ref{sec:conclusions}. 

\section{Setup}
\label{sec:formalism}

\subsection{Sound modes in a superfluid with superflow}

The sound modes can be computed by starting from the hydrodynamic equations, 
\be \label{hydro}
0= \partial_\mu j^\mu  \, , \qquad 0= \partial_\mu s^\mu  \, , \qquad 0= s_\mu(\partial^\mu\Theta^\nu-\partial^\nu\Theta^\mu)  \, .
\ee
In this formulation, the two-fluid nature of the superfluid is obvious: there are two conserved currents, the charge current $j^\mu$ which is conserved due 
to the $U(1)$ symmetry of the underlying microscopic theory, and the entropy current $s^\mu$, which is conserved because dissipation is neglected. 
Each current has an associated conjugate momentum: the conjugate momentum for the charge current can be written as $\partial^\mu\psi$, while the conjugate momentum 
for the entropy current is denoted by $\Theta^\mu$. Here, the scalar field $\psi$ is the phase of the order parameter for superfluidity -- the Bose-Einstein condensate --
and the conjugate momentum is related to the superfluid four-velocity 
\be \label{vmu}
v^\mu = \frac{\partial^\mu\psi}{\sigma} \, , 
\ee
where the Lorentz scalar $\sigma\equiv (\partial_\mu\psi\partial^\mu\psi)^{1/2}$ is identical to the chemical potential measured in the rest frame of the superfluid.
For each of the conjugate momenta there is a vorticity equation. However, the vorticity related to $\partial^\mu\psi$ vanishes trivially, 
$\partial^\mu\partial^\nu\psi - \partial^\nu\partial^\mu\psi = 0$. The third equation in (\ref{hydro}) is the vorticity equation for $\Theta^\mu$,
whose temporal component is the temperature $T$. 

We can express the hydrodynamic equations solely in terms of $\partial^\mu\psi$ and $s^\mu$ with the help of the relations
\be \label{decompose}
j^\mu = \frac{n_n}{s} s^\mu+\frac{n_s}{\sigma}\partial^\mu\psi \, , \qquad \Theta^\mu = -\frac{n_n}{s}\partial^\mu\psi + \frac{w}{s^2}s^\mu \, , 
\ee
where $s$ is the entropy density, and $w\equiv \mu n_n + sT$ the enthalpy density of the normal fluid with the chemical potential $\mu=\partial_0\psi$ 
($s$, $T$, $\mu$ all measured in the normal-fluid rest frame). Moreover, $n_n$ and $n_s$ are the normal-fluid and 
superfluid charge densities, measured in their respective rest frames. They can be computed from the three-current ${\bf j}$,
\be
n_s = - \sigma\frac{\nabla\psi\cdot{\bf j}}{(\nabla\psi)^2} \, , \qquad n_n  = n - \frac{\mu}{\sigma} n_s \, , 
\ee
where $n=j^0$ is the total charge density (measured in the normal-fluid rest frame), and $\mu/\sigma = (1-v^2)^{-1/2}$ is the usual Lorentz factor, with $v\equiv|{\bf v}|$
being the modulus of the superfluid three-velocity ${\bf v} = -\nabla\psi/\mu$. 
The decomposition (\ref{decompose}) translates the formulation in terms of the conserved currents into a formulation in terms of normal-fluid and superfluid components
(whose currents are not separately conserved). 

Sound waves in the linear approximation are small oscillatory deviations from equilibrium. Thus, for instance for the chemical potential, one writes
$\mu({\bf x},t) = \mu + \delta\mu({\bf x},t)$,  where $\mu$ is the equilibrium 
value, and the deviations $\delta\mu({\bf x},t)$ are kept to linear order.  Analogously, the superfluid three-velocity 
is ${\bf v}({\bf x},t) = {\bf v} + \delta{\bf v}({\bf x},t)$, taking into account the non-vanishing equilibrium superflow. In contrast, 
the normal-fluid three-velocity is written as ${\bf v}_n({\bf x},t) = \delta{\bf v}_n({\bf x},t)$ because the equilibrium calculation is performed in the
normal-fluid rest frame. With the help of $\partial_0\psi=\mu$ and the thermodynamic relation
\bea \label{diffP}
dP &\simeq &  nd\mu +sdT -\frac{n_s}{\sigma}\nabla\psi\cdot d\nabla\psi \, ,
\eea
where $P$ is the pressure, one derives two wave equations for the two deviations $\delta\mu({\bf x},t)$ and $\delta T ({\bf x},t)$
from the above hydrodynamic equations. The details of this derivation 
can be found in appendix D of Ref.\ \cite{2013PhRvD..87f5001A}. Assuming harmonic oscillations, one writes 
$\delta\mu({\bf x},t) = \delta\mu_0 e^{i(\omega t - {\bf k}\cdot{\bf x})}$,
$\delta T({\bf x},t) = \delta T_0 e^{i(\omega t - {\bf k}\cdot{\bf x})}$, where $\delta\mu_0$, $\delta T_0$ are the amplitudes and $\omega$ and ${\bf k}$ are energy and
wave vector of the sound wave. Then, the two equations for $\delta\mu_0$, $\delta T_0$ can be written as
\be \label{matrix}
\left(\begin{array}{cc}u^2a_1+[a_2+a_4(\uk\cdot\nabla\psi)^2]+u\,a_3\uk\cdot\nabla\psi & u^2b_1+b_2+u\,b_3\uk\cdot\nabla\psi \\ [2ex]
u^2A_1+[A_2+A_4(\uk\cdot\nabla\psi)^2]+u\,A_3\uk\cdot\nabla\psi & u^2B_1+B_2+u\,B_3\uk\cdot\nabla\psi \end{array}\right)\left(\begin{array}{c}
\delta\mu_0 \\[2ex] \delta T_0 \end{array}\right) = 0  \, ,
\ee
where $u=\omega/k$ is the sound speed. 
The various coefficients are functions of $T$, $\mu$, and $|\nabla\psi|$ and are evaluated in equilibrium. 
In the presence of a superflow they are complicated; their explicit form is given in appendix \ref{appA}. 
Nontrivial solutions for $\delta\mu_0$, $\delta T_0$ require the determinant of the above $2\times 2$ matrix to vanish. 
This condition yields a quartic polynomial for the sound speed of the form 
\bea \label{poly}
0&=&u^4Q^{(4)}+u^3Q^{(3)}\uk\cdot\nabla\psi+u^2\Big[Q^{(2)}_{1}+Q^{(2)}_{2}(\uk\cdot\nabla\psi)^2\Big] \non[2ex]
&&+u\,\Big[Q^{(1)}_{1}+Q^{(1)}_{2}(\uk\cdot\nabla\psi)^2\Big]\uk\cdot\nabla\psi+\Big[Q^{(0)}_{1}+Q^{(0)}_{2}
(\uk\cdot\nabla\psi)^2\Big] \, , 
\eea
where all angular dependence is written explicitly. Again, the coefficients of this polynomial can be found in 
appendix \ref{appA}. Obviously, the sound velocities will depend on the angle $\theta$ between the superfluid velocity 
${\bf v}\propto -\nabla\psi$ and the direction of the sound
wave $\uk$. If $u({\bf k})$ is a solution, Eq.\ (\ref{poly}) shows that $-u(-{\bf k})$ is also a solution. Consequently, if there are two solutions $u_1({\bf k})$
 and $u_2({\bf k})$ that are positive for all angles, there will be two corresponding negative
solutions for any angle that can be discarded. In general, the solutions can become complex. If $u$ is a solution, the complex conjugate $u^*$ is also 
a solution because all coefficients of the polynomial are real. Therefore, in that case, $\delta\mu_0, \delta T_0\propto e^{-\gamma t}$, where $\gamma = k\,{\rm Im}(u)$
assumes a positive value for one sound mode and the negative value with the same magnitude for the other sound mode: one mode decays, one mode wants to explode. 
This is exactly the kind of instability that has been discussed in a general two-fluid system in Ref.\ \cite{Samuelsson:2009up}, where it has been identified with 
the two-stream instability known from plasma physics.  

In order to analyze the instability it is also useful to discuss the amplitudes of the sound waves. The ratio of the amplitudes in chemical potential
and temperature can obviously be computed through 
\be
\frac{\delta T_0}{\delta\mu_0} = -\frac{u^2a_1+[a_2+a_4(\uk\cdot\nabla\psi)^2]+u\,a_3\uk\cdot\nabla\psi}{u^2b_1+b_2+u\,b_3\uk\cdot\nabla\psi} 
\, , 
\ee
where $u$ is a solution of Eq.\ (\ref{poly}).
It is useful to define the mixing angle \cite{Alford:2013koa}
\be \label{alpha}
\alpha \equiv \arctan\frac{\delta T_0}{\delta \mu_0} \, . 
\ee
If $u\in \mathbb{R}$, then also $\alpha\in\mathbb{R}$, and the mixing angle 
says whether a given sound mode is a pure chemical potential wave ($\alpha=0$) or a pure temperature wave ($|\alpha|=\frac{\pi}{2}$) 
or some mixture of both. The sign of $\alpha$ determines whether chemical potential and temperature oscillate in phase ($\alpha>0$) or out of phase ($\alpha<0$).
Complex values of $u$ lead to a non-trivial phase factor between chemical potential and temperature oscillations.

\begin{figure}[t] 
\begin{center}
\hbox{\includegraphics[width=0.5\textwidth]{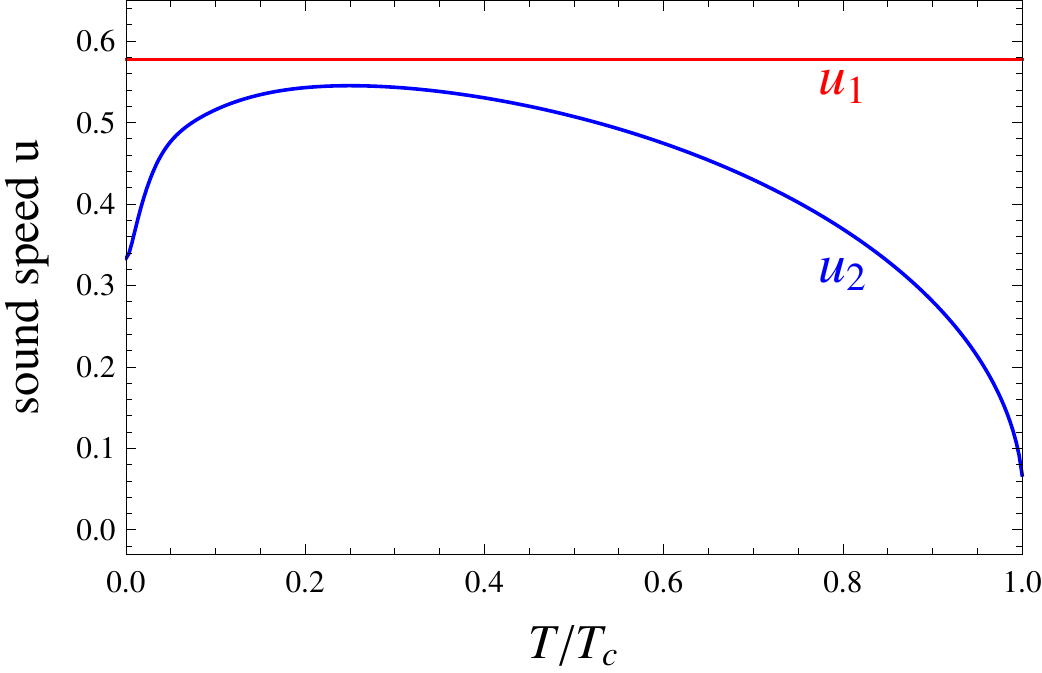}\includegraphics[width=0.5\textwidth]{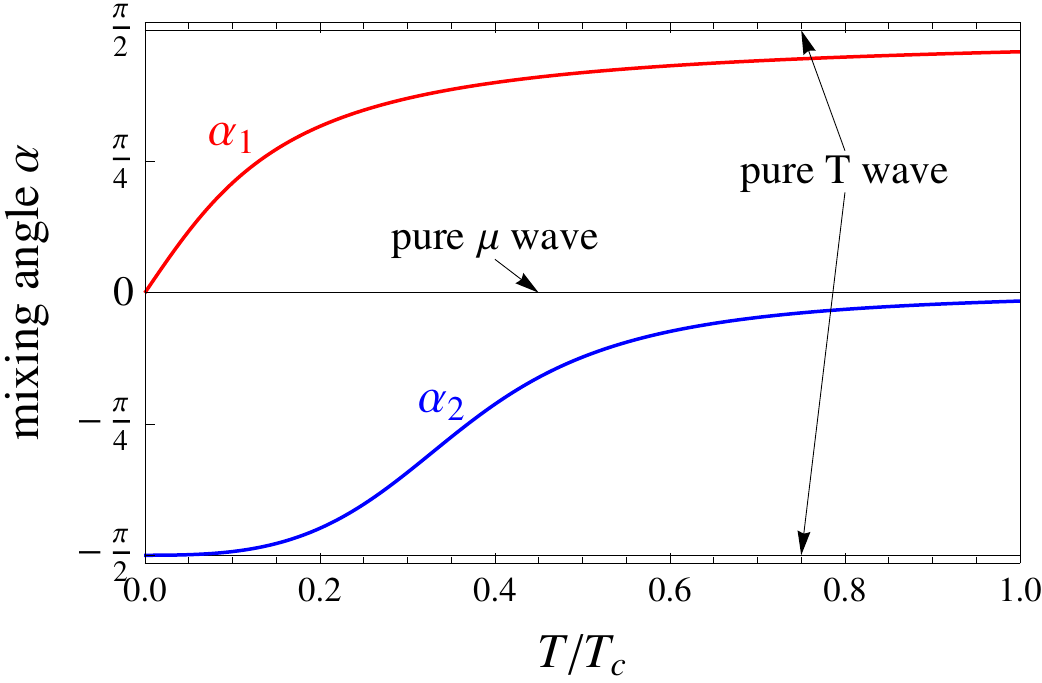}}
\caption{Left panel: speeds of first and second sound as a function of temperature in the absence of a superflow, ${\bf v}=0$, where there are no instabilities,
i.e., $u_1,u_2\in\mathbb{R}$. 
Right panel: corresponding mixing angles, showing that first and second sound reverse their roles, the first (second) sound evolving from a pure $\mu$ (pure $T$)
wave at small temperatures to an almost pure $T$ (pure $\mu$) wave at large temperatures. The results are obtained with the parameters $m=0$ and $\lambda=0.05$
(since the shown quantities are all dimensionless, the chemical potential $\mu$ drops out and does not have to be specified). 
In the ultra-relativistic limit without superflow, $m={\bf v}=0$, one finds $u_1=1/\sqrt{3}$ for all temperatures, and the mixing angles become particularly simple, 
$\alpha_1=\arctan(T/\mu)$, $\alpha_2=-\arctan(n/s)$ \cite{Alford:2013koa}.  
}
\label{fig:v0}
\end{center}
\end{figure}

The two sound speeds for vanishing superflow as a function of temperature are shown in Fig.\ \ref{fig:v0} \cite{Alford:2013koa}. In this case,  
$u_1,u_2\in \mathbb{R}$, and no 
instability occurs. This result is obtained within the microscopic model that I will now describe.

\subsection{Microscopic model and self-consistent formalism}

I use the same model and formalism as in Refs.\ \cite{2013PhRvD..87f5001A,Alford:2013koa}. All details can be found in these references, and here I only briefly
summarize the ingredients of the calculation. The starting point is the following $U(1)$ symmetric Lagrangian for a complex scalar field $\varphi$,
\be \label{L}
{\cal L} = \partial_\mu\varphi\partial^\mu\varphi^* - m^2 |\varphi|^2 - \lambda |\varphi|^4 \, , 
\ee
with the boson mass $m>0$ and the coupling constant $\lambda>0$. Superfluidity occurs through a Bose-Einstein condensate $\langle\varphi\rangle$. 
To this end, one needs to introduce a chemical potential $\mu>m$. And, to investigate the hydrodynamics of the superfluid, one has to make the condensate move, 
i.e., introduce a superfluid velocity ${\bf v}$. Both is done with the help of the phase of the condensate, $\langle\varphi\rangle = \rho e^{i\psi}$, with $\psi$ being
the scalar field introduced above, see Eq.\ (\ref{vmu}). The superfluid three-velocity ${\bf v}$ and the modulus of the condensate $\rho$
are assumed to be constant in space and time.  Together with the temperature $T$, $\mu$ and ${\bf v}$ are the externally given parameters. They are all measured in the 
rest frame of the normal fluid, in which the field-theoretical calculation is performed. Employing this uniform ansatz and taking the 
dissipationless limit are important assumptions because they significantly simplify the calculation. Nevertheless, for possible resolutions,  damping, or  
time evolution of the two-stream instability discussed later, releasing one or both of these assumptions will be interesting extensions of the present work for the future.

As shown in the previous subsection, the microscopic calculation needs to provide the thermodynamic equilibrium properties of the system.
They are computed as follows. I start from 
the 2PI effective action, truncated at two-loop order, and work in the Hartree approximation in which the self-energy is momentum-independent. 
Minimizing the effective action, one obtains self-consistent equations for the condensate and the boson propagator. 
%The propagator 
%can be parametrized in terms of two mass parameters, and eventually one obtains three coupled equations that can be solved numerically. 
The result of these equations can be re-inserted into the effective action to obtain the pressure, and thus, via taking derivatives of the pressure, all 
thermodynamic quantities that are needed. 

There are various theoretical obstacles in this approach. Firstly, in the given truncation, the Goldstone theorem is violated 
\cite{Baym:1977qb,AmelinoCamelia:1997dd,Andersen:2008tn}. Therefore, I implement the existence of an exactly massless mode by a modification of the 
stationarity equations, keeping the effective action unchanged. As a consequence, the effective action is evaluated not at the minimum, but at a point away 
from that minimum. This particular modification to the stationarity equations, explained in detail in Ref.\ \cite{Alford:2013koa}, 
was already used in Ref.\ \cite{Alford:2007qa}; a similar, but not identical, modification has been suggested in Ref.\ \cite{Pilaftsis:2013xna}, where the minimum 
in the subspace constrained by the requirement of a massless mode is determined. 

Secondly, the renormalization of the approach is non-trivial. This is due to the resummation of a certain class
of diagrams that contains all orders in the coupling constant, and due to the use of the Hartree approximation. I follow the renormalization procedure that has been 
developed in the literature for this approach \cite{Blaizot:2003br,Blaizot:2003an,Andersen:2006ys,Fejos:2007ec,Andersen:2008tn}. However, the presence of a nonzero
superflow further complicates this procedure because the pressure shows an 
ultraviolet divergence that depends explicitly on the superflow \cite{Alford:2013koa}. Therefore, the usual vacuum subtraction, corresponding to a standard 
renormalization condition, cannot 
be applied. This problem apparently induces an ambiguity in the dependence of physical quantities on the superflow. 
I will not attempt to solve this problem in this paper, but rather restrict myself to small values of the coupling strength where this problem is not relevant.
The reason is that at sufficiently small values of the coupling constant, the subleading 
terms that are sensitive to the renormalization procedure (and depend on the renormalization scale) can be neglected \cite{Alford:2013koa}. 

Thirdly, in the Hartree approximation, the transition to the 
non-superfluid phase (at vanishing superflow) is of first order, while a more complete treatment shows a second-order phase transition 
\cite{Baacke:2002pi,Marko:2012wc,Marko:2013lxa}. This problem is also circumvented by the restriction to the weakly coupled regime because then 
the discontinuity of the condensate at the critical point is very small. Nevertheless, as one can see in Fig.\ \ref{fig:v0}, the speed of second sound does not
exactly go to zero at the critical temperature, as it should in a second-order phase transition.

\section{Sound mode instabilities}
\label{sec:twostream}

\subsection{Zero-temperature limit}
\label{sec:zeroT}

For infinitesimally small temperatures, $T\to 0$, the two
sound velocities can be computed analytically in the weak-coupling approximation. (At exactly $T=0$ there is only one fluid and thus only one sound mode.)
Setting the boson mass to zero, $m=0$, i.e., working in the ultra-relativistic
limit, the sound velocities become \cite{2013PhRvD..87f5001A}
\begin{subequations}\label{u12}
\bea \label{u1}
u_1(T\to 0) &=& \frac{\sqrt{3-v^2(1+2\cos^2\theta)}\sqrt{1-v^2}+2v\cos\theta}{3-v^2} \, ,  \\[2ex]
u_2(T\to 0) &=& \frac{\sqrt{9(1-v^2)(1-3v^2)+v^2\cos^2\theta}+v\cos\theta}{9(1-v^2)}  \, . 
\eea
\end{subequations}
These expressions show that there is a critical point at $v=\frac{1}{\sqrt{3}}$, where the speed of first sound becomes negative for $\theta = \pi$,
and the speed of second sound becomes complex for $\theta=\frac{\pi}{2}$. 

In general, Landau's critical velocity is given by the point where the quasiparticle excitations, in the presence of a superflow, become negative,
\be \label{e0}
\epsilon_{\bf k}({\bf v})<0 \, .
\ee
In the formalism used here, $\epsilon_{\bf k}$ is computed from the poles of the self-consistently determined propagator. The relevant pole corresponds to the 
Goldstone mode (while there is also a massive mode which is negligible at low temperatures). 
%Due to the self-consistent treatment, the dispersion is not simply obtained by a 
%Lorentz transformation of the quasiparticle excitations in the absence of a superflow, but it also contains the backreaction of the condensate.
In the zero-temperature, weak-coupling limit, the low-energy dispersion of the Goldstone mode is given by the speed of first sound, $\epsilon_{\bf k}({\bf v}) = u_1 k$. 
Therefore, in this limit, Landau's critical velocity and the velocity at which the speed of second sound becomes complex, are identical. 
The following results will show that this is no longer true at nonzero temperatures. There, (\ref{e0}) sets in for larger values of the superfluid velocity than the
one at which a sound mode becomes complex, opening up a window for the two-stream instability.

\subsection{Upstream direction}
\label{sec:nonzeroT}

\begin{figure}[t] 
\begin{center}
\hbox{\includegraphics[width=0.48\textwidth]{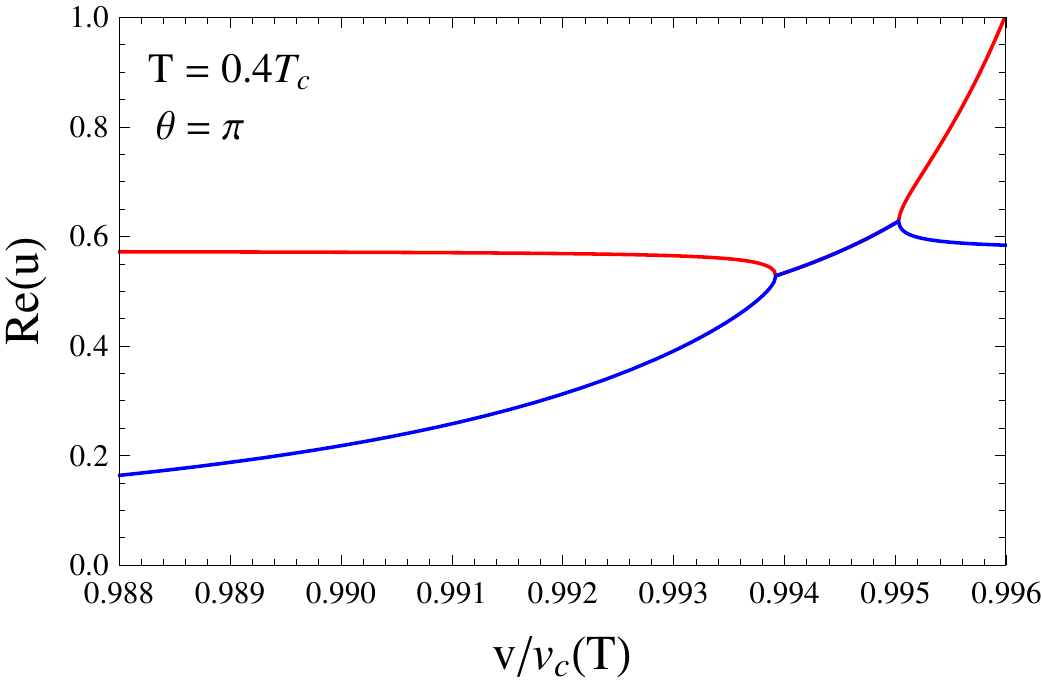}\includegraphics[width=0.5\textwidth]{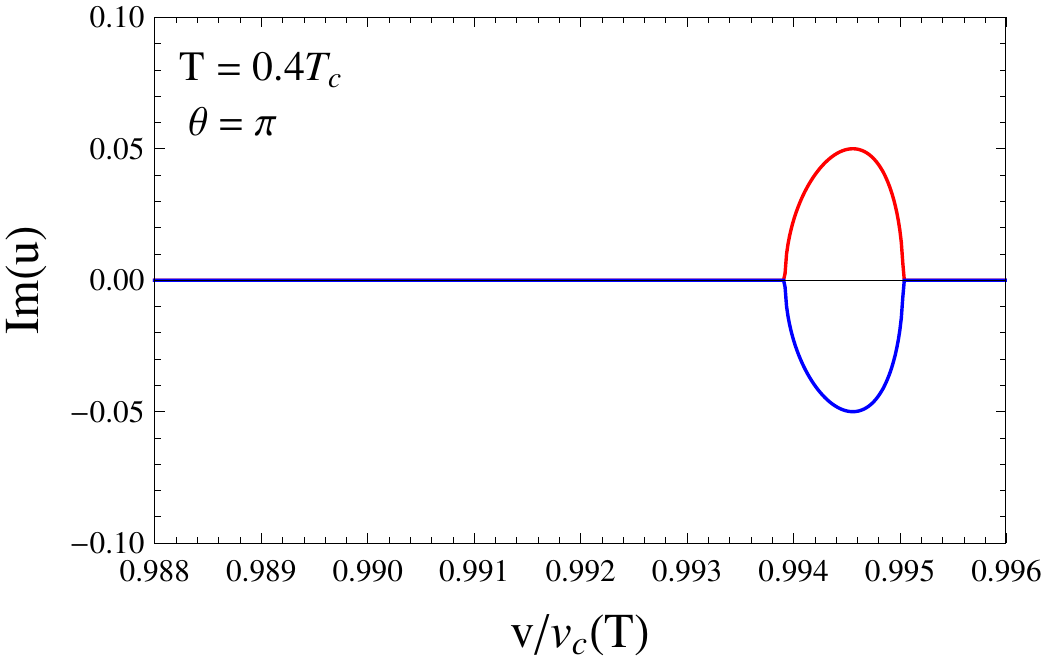}}
\caption{Real parts (left) and imaginary parts (right) of the two sound speeds for sound waves propagating opposite to the superflow, $\theta=\pi$, at a temperature 
  $T=0.4\,T_c$ with $T_c$ being the critical temperature in the absence of a superflow. The superflow is a background relative flow between superfluid and normal 
fluid on top of which the sound modes propagate. The parameters of the model are chosen as $m=0$, $\lambda=0.05$.
The superfluid velocity $v$ is given in units of $v_c(T)$, the critical velocity beyond which the Goldstone dispersion becomes negative (Landau's critical velocity). 
Instabilities in the form of nonzero imaginary parts of the sound modes set in slightly below $v_c(T)$. The horizontal scale ends 
at the point beyond which there is a mode with $u>1$. }
\label{fig:u1u2}
\end{center}
\end{figure}

At nonzero temperatures, the self-consistent calculation does not allow for simple analytical results, and one has to proceed numerically. 
The derivatives of the pressure that are needed for the sound wave equation are all computed in a semi-analytical way, by taking the derivatives of the various integrands
analytically and then performing the three-momentum integrals numerically. Importantly, this reduces the possible sources for numerical errors tremendously 
because there is no need to work with finite differences or any other numerical methods to compute derivatives. I exactly follow the calculation 
of Ref.\ \cite{Alford:2013koa}, where more details can be found.

\begin{figure}[t] 
\begin{center}
\hbox{\includegraphics[width=0.5\textwidth]{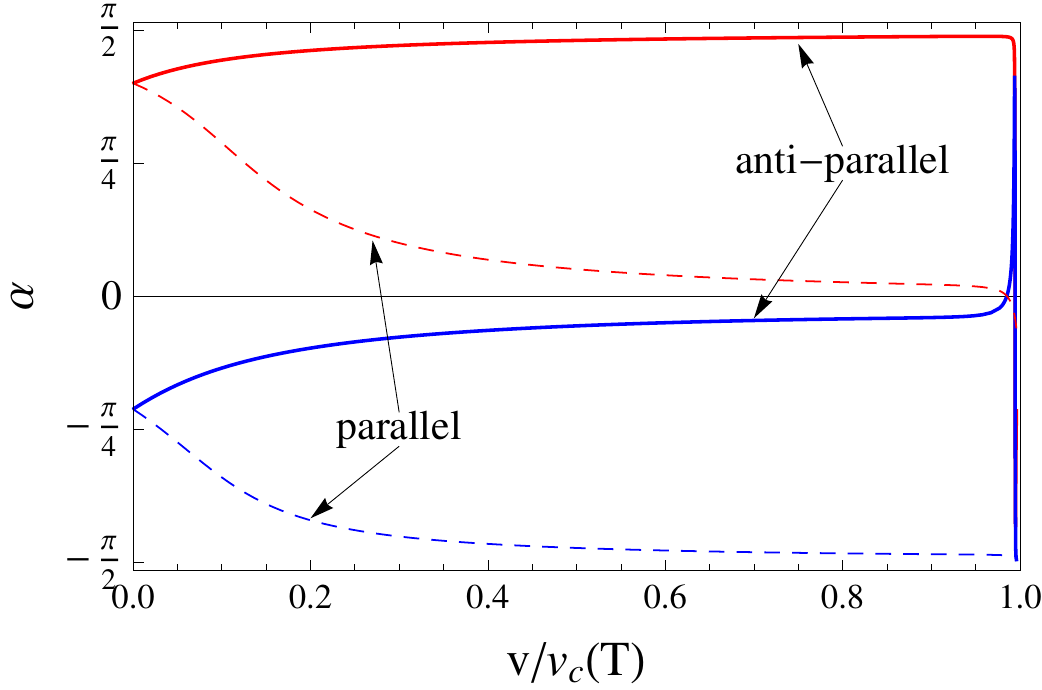}\includegraphics[width=0.49\textwidth]{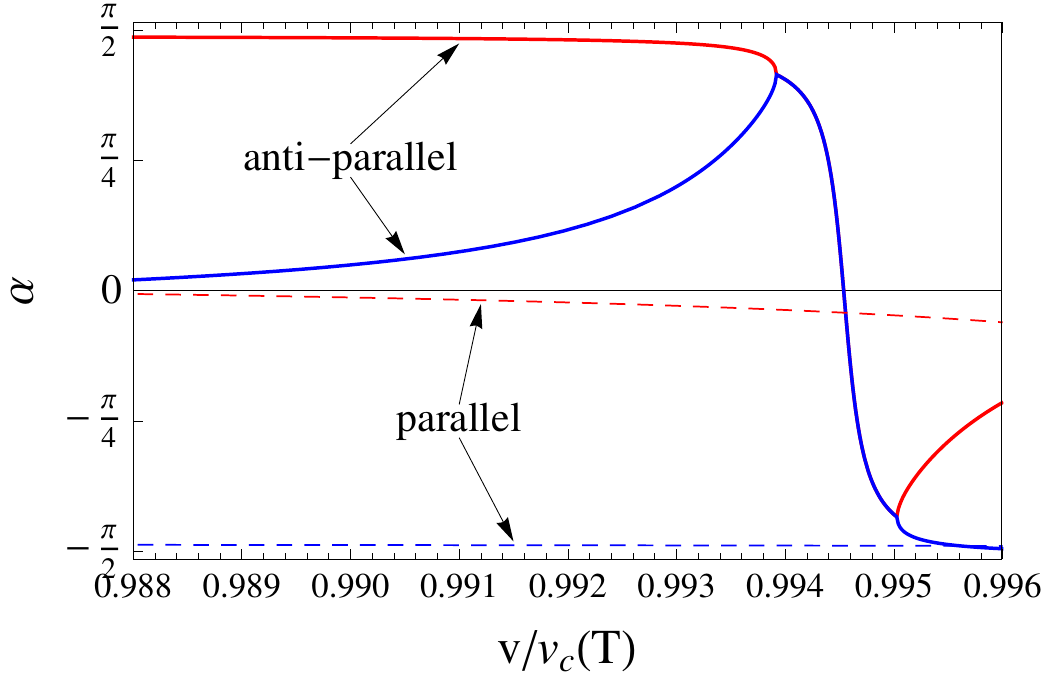}}
\caption{Mixing angle $\alpha$ for the sound modes shown in Fig.\ \ref{fig:u1u2}, i.e., for $T=0.4\,T_c$. In addition to the modes propagating anti-parallel 
to the superflow ($\theta=\pi$, solid lines), the mixing angle is also plotted for the sound modes parallel to the superflow ($\theta=0$, thin dashed lines).
The left panel shows $\alpha$ for all superfluid velocities from $v=0$ up to Landau's critical velocity $v_c(T)$; the right panel zooms into the unstable region
close to $v_c(T)$. In the unstable region, where the real parts of the sound velocities coincide, the plot shows $\arctan[{\rm Re}\,(\delta\mu_0/\delta T_0)]$.} 
\label{fig:alpha}
\end{center}
\end{figure}

A typical manifestation of the instability is shown in Fig.\ \ref{fig:u1u2}. This figure shows real and imaginary parts of the two sound speeds as a function of the 
superflow, for a fixed temperature $T=0.4\,T_c$, where $T_c$ is the critical temperature in the absence of a superflow. The parameters of the 
Lagrangian are chosen to be $m=0$ and $\lambda=0.05$, which corresponds to the ultra-relativistic and weak-coupling limits. I will work with these values throughout the 
paper except for Sec.\ \ref{sec:dependence}, where the dependence of the results on $m$ and $\lambda$ is discussed. The critical velocity $v_c(T)$ 
is determined numerically by computing the onset of negative quasiparticle energies. The plots show the sound speeds below, but very close
to, that critical velocity, and focus on a single direction of the sound wave, namely anti-parallel to the superflow (``upstream''). 
In the chosen temperature regime, this is the most interesting direction because the instabilities occur in this direction first (= for the lowest superfluid
velocities).

One can see that there is a certain superfluid velocity at which both sound speeds become complex, their values being complex conjugate
to each other, as discussed above. The plots are qualitatively similar to Fig.\ 1 in Ref.\ \cite{Samuelsson:2009up}, where the sound modes have been 
computed for a general two-fluid system and show the same instability in a certain parameter regime. In that reference, certain parameters such as the 
entrainment coefficient are varied by hand, while here they become functions of $T$, $\mu$, and ${\bf v}$, once the parameters $m$ and $\lambda$ of the microscopic 
theory are fixed. Following Ref.\ \cite{Samuelsson:2009up}, the instability seen here can be identified with the two-stream instability which is known from 
plasma physics and which applies to superfluids because of their two-fluid nature \cite{2004MNRAS.354..101A}. 

It is crucial that there is a coupling 
between the two fluids in order for the two-stream instability to occur. Such a coupling can be characterized by the entrainment coefficient that relates the two
currents $j^\mu$ and $s^\mu$ to the conjugate momentum of the other current. This entrainment coefficient has been computed in the present model,
and it was found that, in the absence of a superflow, it vanishes for zero temperature and increases monotonically with increasing temperature \cite{Alford:2013koa}. 
This is in accordance with the absence of the instability at zero temperature. However, the coupling between the two fluids that is relevant for the sound modes 
is not only given by the entrainment coefficient: a covariant 
two-fluid description can be formulated in terms of a generalized energy density $\Lambda$ that depends on the Lorentz scalars $j^2$, $s^2$, and $j\cdot s$. 
A term proportional to $j\cdot s$ gives rise to a nonzero entrainment coefficient, which is given by the first derivative 
$\frac{\partial\Lambda}{\partial(j\cdot s)}$. A coupling is also induced by a term proportional to $j^2s^2$, giving rise to a nonzero second derivative 
$\frac{\partial^2\Lambda}{\partial j^2 \partial s^2}$. Both kinds of couplings enhance the two-stream instability \cite{Samuelsson:2009up}.

Fig.\ \ref{fig:u1u2} shows that the instability appears to vanish again for sufficiently large superfluid velocities. However, as will be 
discussed in Sec.\ \ref{sec:angles}, 
sound modes in other directions (other than $\theta=\pi$) become unstable too, and the instability persists up to larger superfluid velocities than the 
result for the upstream direction suggests. Eventually, going even closer to the critical velocity $v_c(T)$, the sound speed of one of the 
sound modes becomes larger than one, i.e., larger than the speed of light. The horizontal axis in Fig.\ \ref{fig:u1u2} stops at the point where this happens.
(Beyond the given scale, there is a divergence of the sound speed at $v/v_c(T)\simeq 0.9984$.) I have checked that for all superfluid velocities smaller than $v_c(T)$,
including the regime where there is a sound speed larger than one, all thermodynamic quantities such as entropy, superfluid density etc.\ behave regularly.
It is thus not clear whether this curious behavior indicates another physical instability or a problem with the calculation. Due to the semi-analytical 
evaluation of the thermodynamic functions described above, the numerics are very stable and a numerical error as a source for this behavior is thus very unlikely. 
As mentioned
above, however, there are several approximations used in the appraoch such as the Hartree approximation and the way the Goldstone theorem is implemented. 
It would be therefore be interesting to see whether the same observations are made after going beyond these approximations or in a completely different 
microscopic model. 

Some better understanding of the two-stream instability can be gained by computing the mixing angle $\alpha$ 
defined in Eq.\ (\ref{alpha}). Its value for the two sound modes is shown in 
Fig.\ \ref{fig:alpha}. Before reaching the unstable regime, first and second sound can be distinguished by the sign of $\alpha$, i.e., the first sound is an in-phase
oscillation of $\mu$ and $T$, while the second sound is an out-of-phase oscillation. Just before the critical regime, first sound is, for this particular 
temperature, almost a pure temperature oscillation, while second sound is predominantly an oscillation in chemical potential. At some point, just before the 
instability sets in, the second sound undergoes a dramatic change: it turns from a predominant $\mu$, out-of phase oscillation into a predominant $T$, in-phase
oscillation and thus becomes indistinguishable from first sound. This is the point where the instability sets in.

\begin{figure}[t] 
\begin{center}
\hbox{\includegraphics[width=0.33\textwidth]{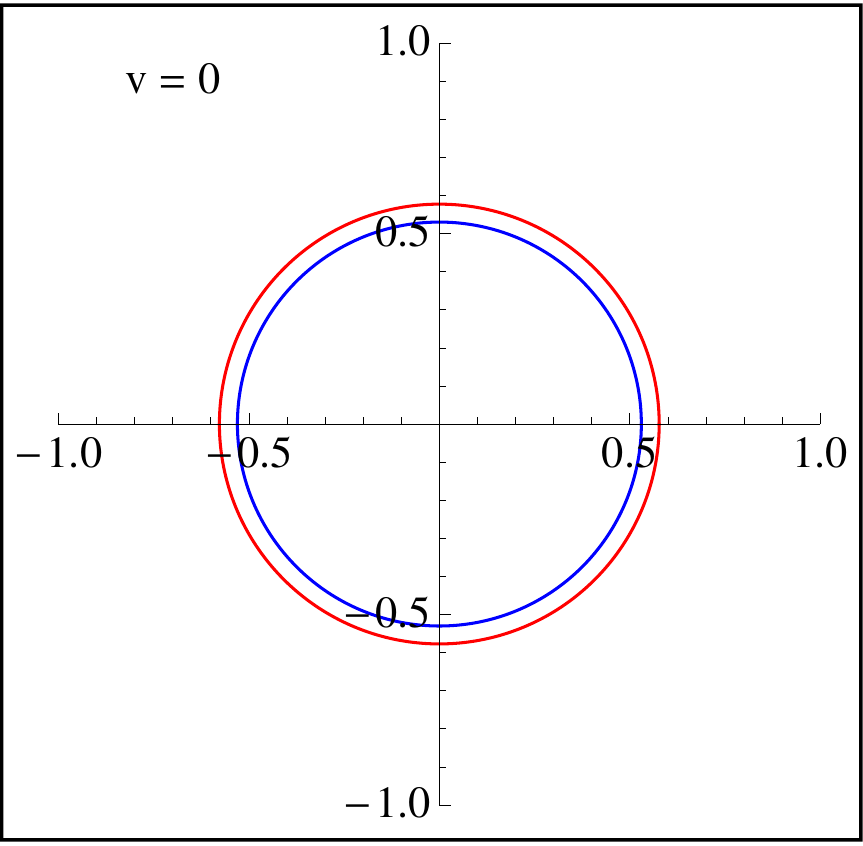}\hspace{0.05cm}\includegraphics[width=0.33\textwidth]{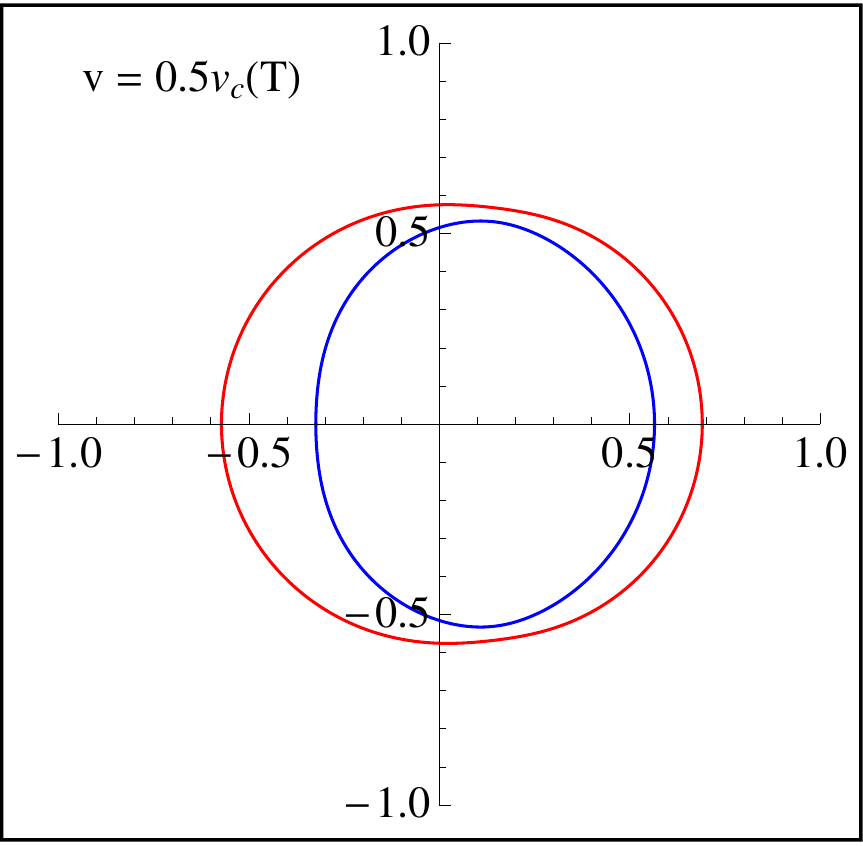}\hspace{0.05cm}\includegraphics[width=0.33\textwidth]{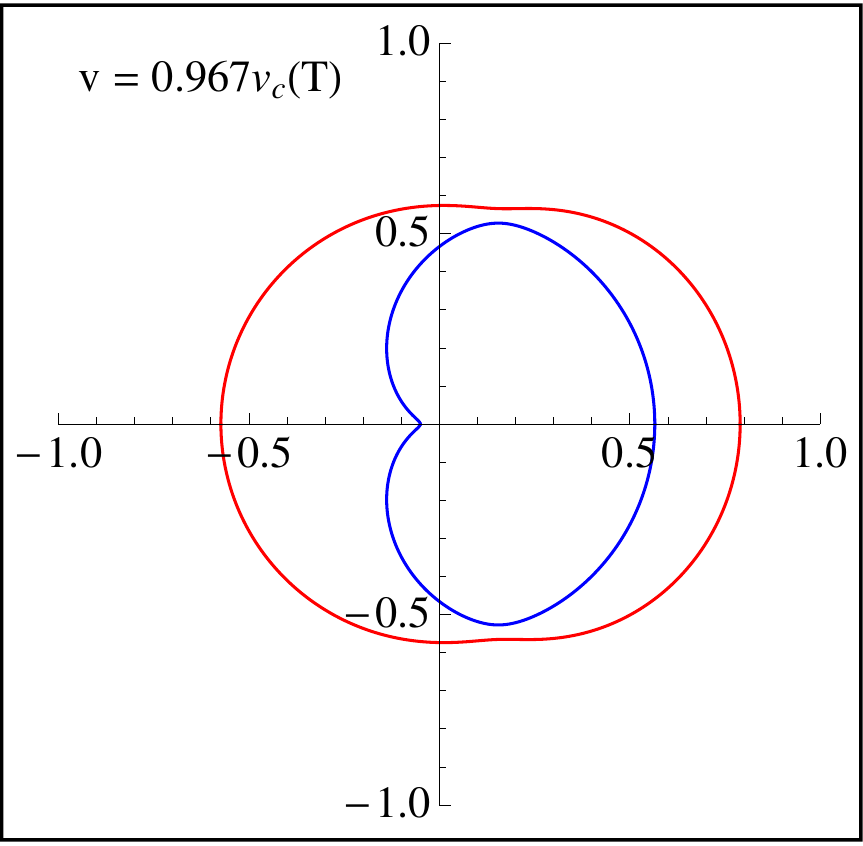}}

\hbox{\includegraphics[width=0.33\textwidth]{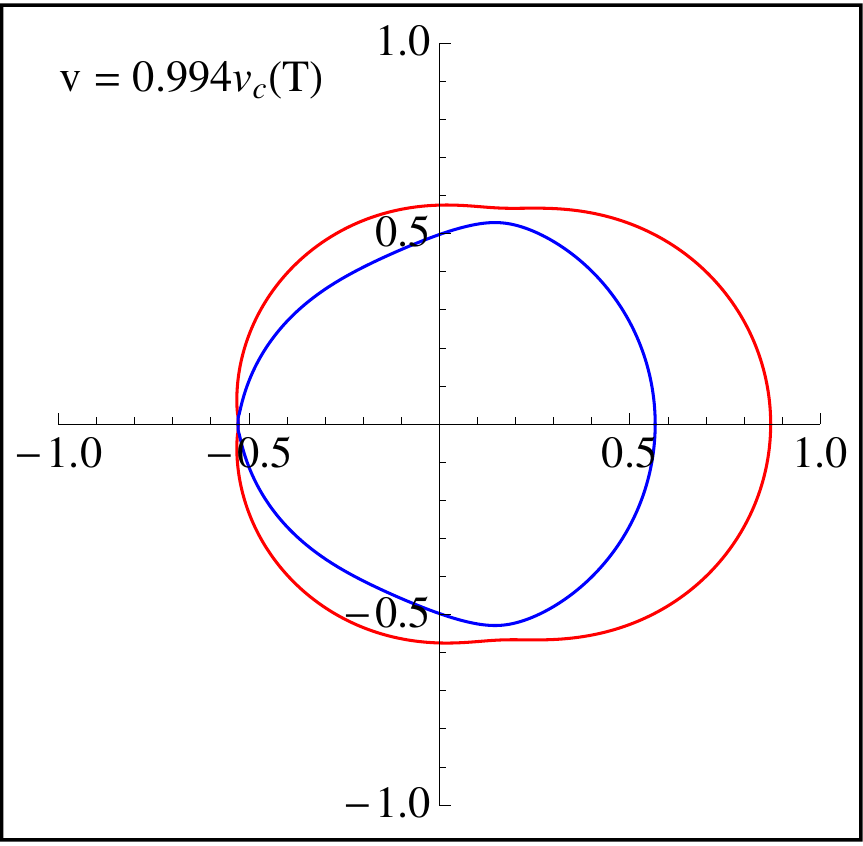}\hspace{0.05cm}\includegraphics[width=0.33\textwidth]{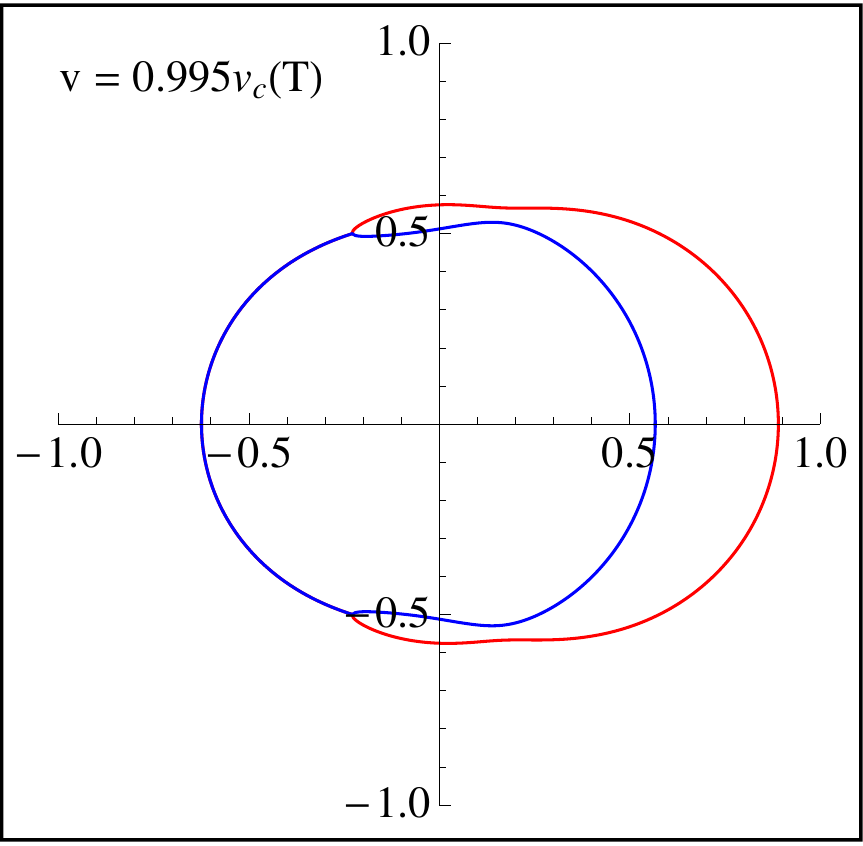}\hspace{0.05cm}\includegraphics[width=0.33\textwidth]{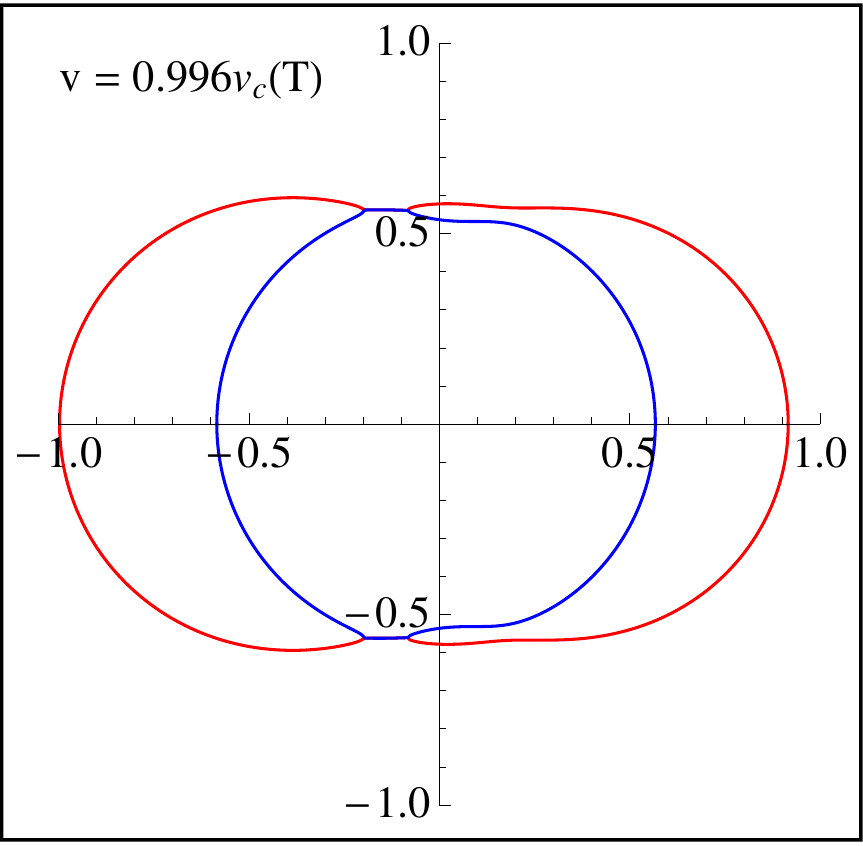}}
\caption{Real part of the sound velocities for $T=0.4T_c$ and 6 different values of the superfluid velocity $v$. The sound speed for a given angle 
is given by the distance of the curve from the origin. The superfluid velocity ${\bf v}$ points to the right. The first row 
shows stable configurations where the superflow ``drags'' the sound modes to the right. In the second row there is always an angular regime where 
$u_1$ and $u_2$ have the same real part and thus must both have 
imaginary parts, equal in magnitude, but opposite in sign. The last plot shows the point where one of the sound speeds reaches the speed of light.
}
\label{fig:polar1}
\end{center}
\end{figure}

\subsection{All directions}
\label{sec:angles}

\begin{figure}[t] 
\begin{center}
\hbox{\includegraphics[width=0.33\textwidth]{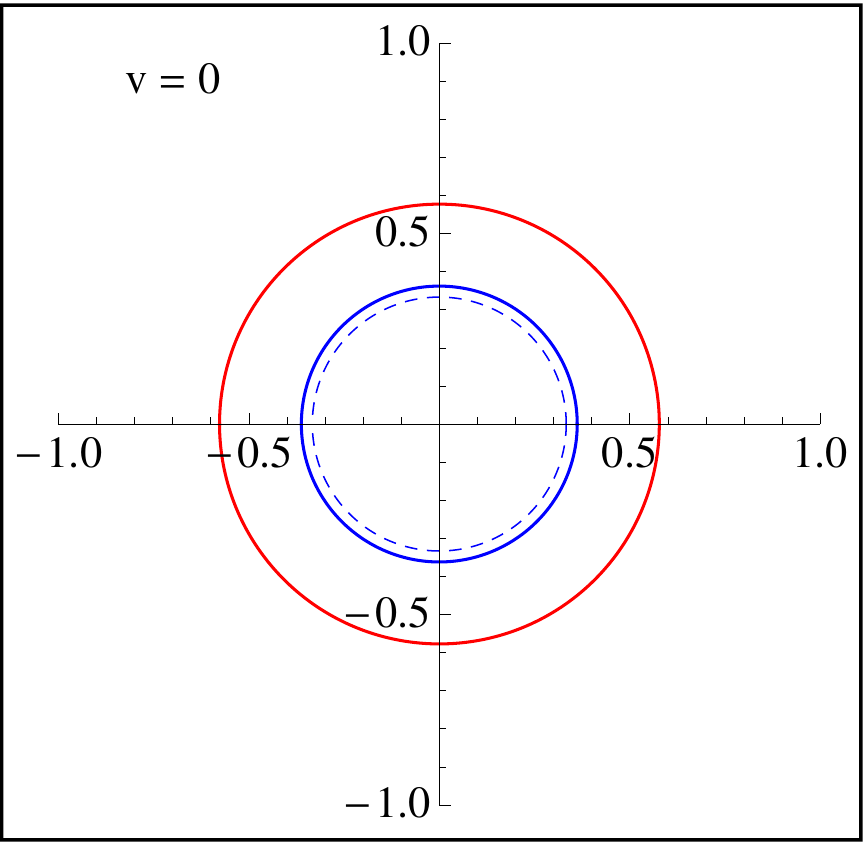}\hspace{0.05cm}\includegraphics[width=0.33\textwidth]{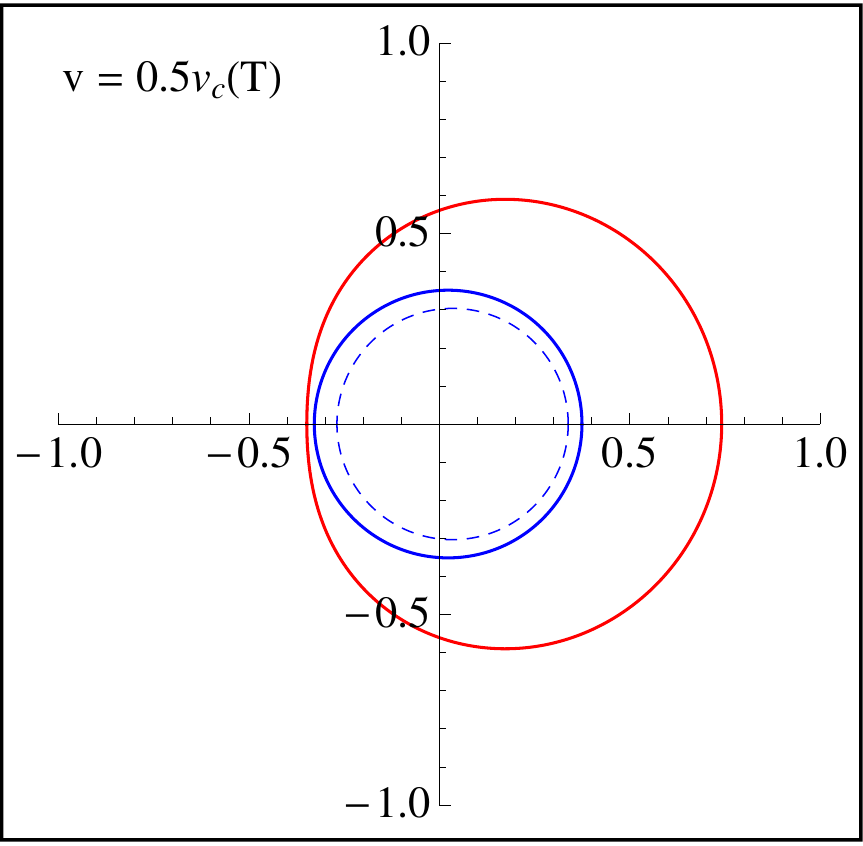}\hspace{0.05cm}\includegraphics[width=0.33\textwidth]{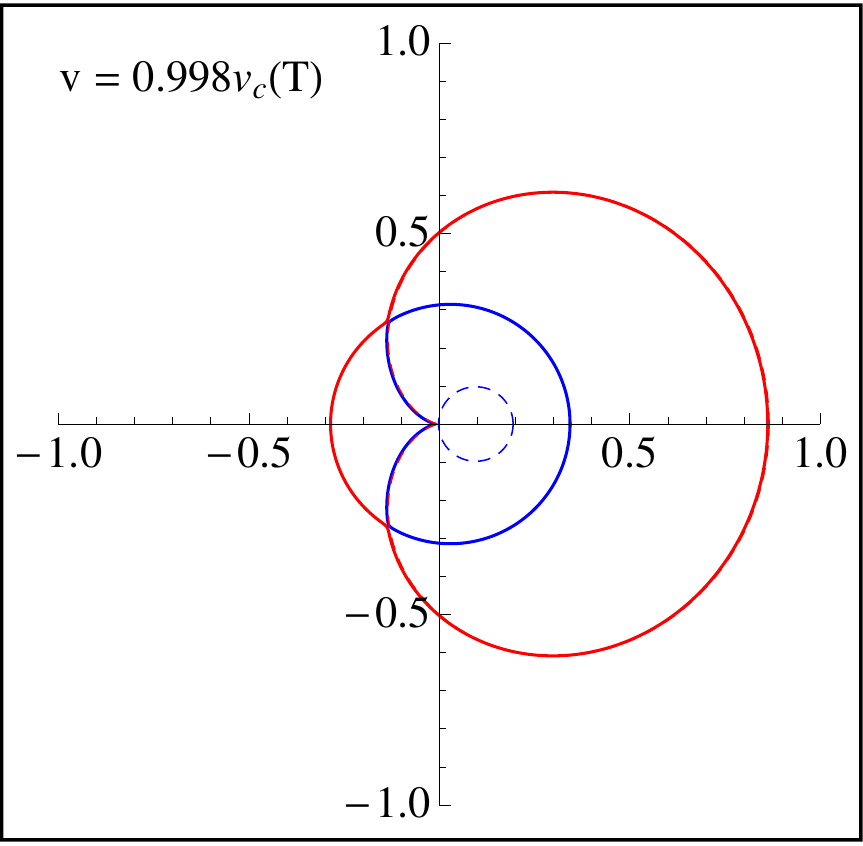}}

\hbox{\includegraphics[width=0.33\textwidth]{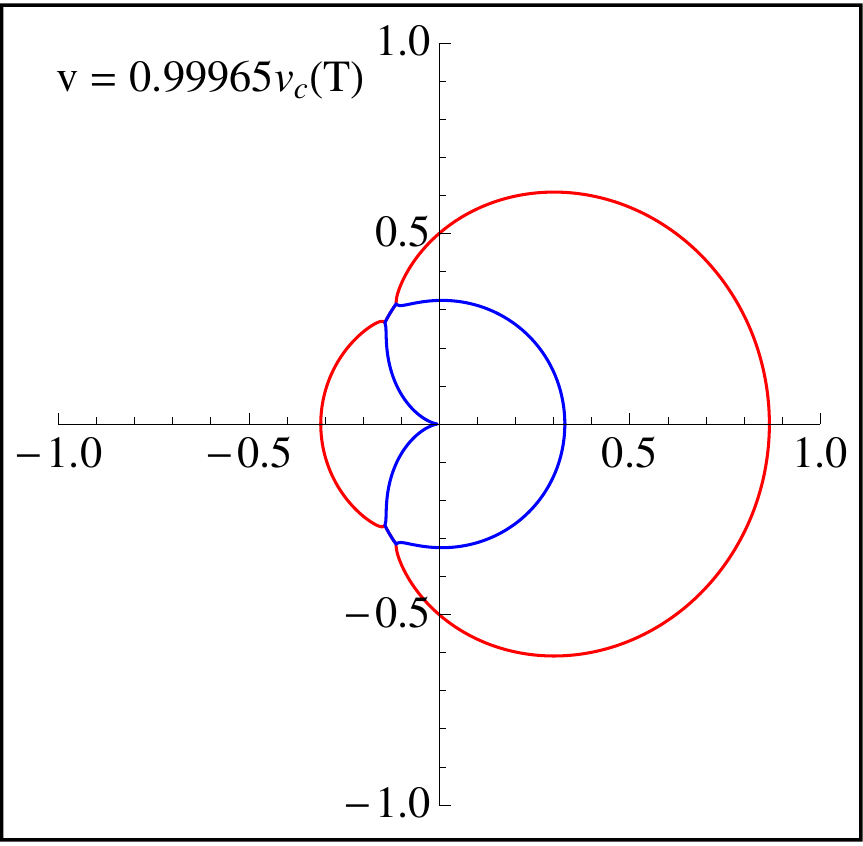}\hspace{0.05cm}\includegraphics[width=0.33\textwidth]{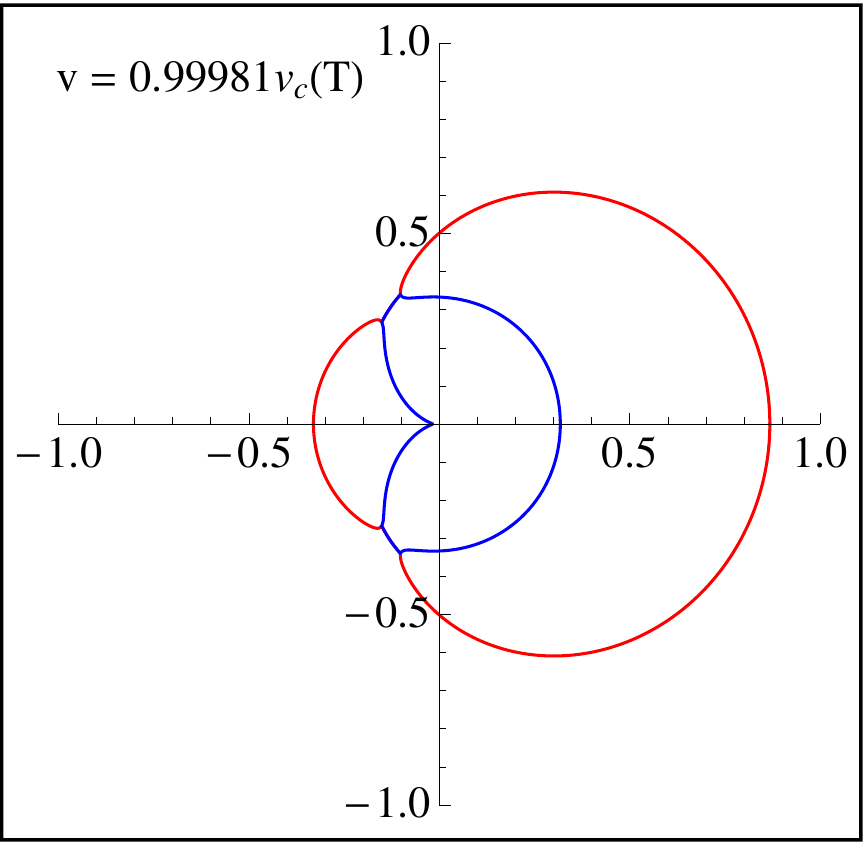}\hspace{0.05cm}\includegraphics[width=0.33\textwidth]{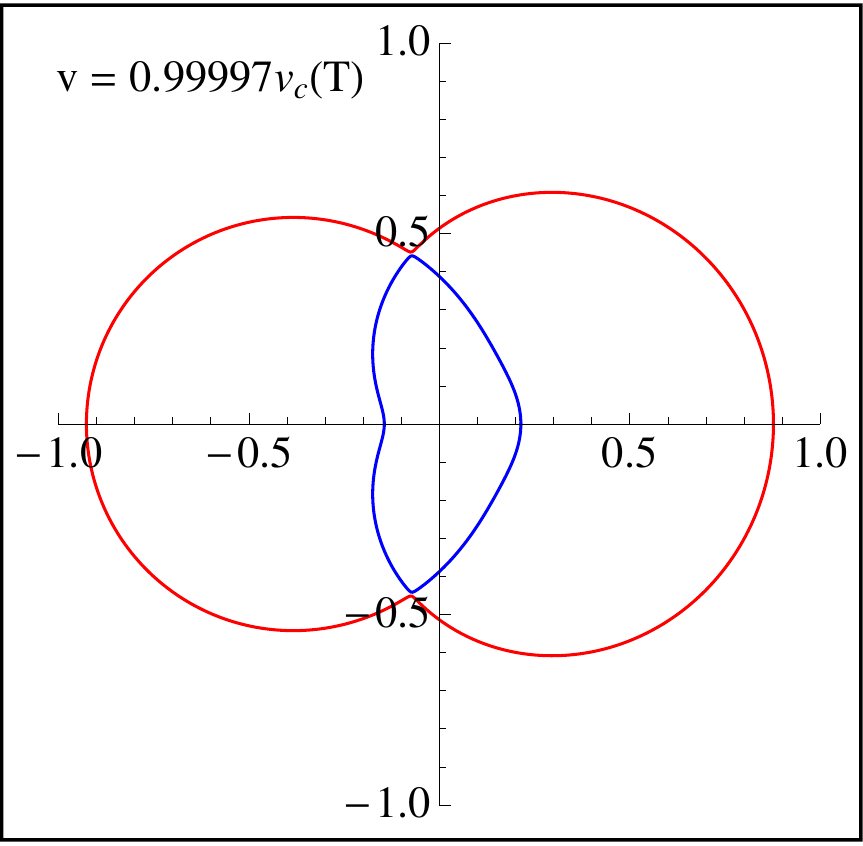}}

\caption{Real parts of the sound speeds, as in Fig.\ \ref{fig:polar1}, but for a very small temperature $T=0.009\,T_c$. The qualitative difference to intermediate
and large temperatures is that the sound modes anti-parallel to the superflow do not acquire an imaginary part, the instability rather occurs only at non-trivial angles
(where the real parts of the two modes become identical). The dashed curve in the upper panels is the analytical $T\to 0$ result for $u_2$ from Eq.\ (\ref{u12}).
The $T\to 0$ result for $u_1$ is indistinguishable from the numerical result. }
\label{fig:polar2}
\end{center}
\end{figure}

A study of the full angular dependence shows that the upstream direction is not the only direction
where the two-stream instability occurs. 
In Fig.\ \ref{fig:polar1}, the sound velocities are plotted for all directions. The temperature is the same as above, $T=0.4\, T_c$, and
each panel corresponds to one value of the superfluid velocity. The picture that emerges is as follows: as the superflow is increased, the downstream velocities
increase while the upstream velocities decrease, as one might expect. This effect is barely visible for the first sound, and more pronounced for the second sound. 
In particular, the second sound speed for sound propagation exactly opposite to the superflow ``wants'' to become zero as $v$ is getting larger. 
Since the calculation is performed in the normal-fluid rest frame, this means that the second sound ``wants'' to move together with the normal fluid. 
Eventually, one might expect that the superflow is strong enough to ``drag'' both the downstream and the (initially) upstream 
modes into the same direction, seen from the normal fluid rest frame. In other words, in this scenario, there would exist a sound mode that, seen from the normal fluid 
rest frame, propagates in the opposite direction as seen from the superfluid rest frame. This scenario never occurs. Maybe one way to interpret the instability is to 
say that it prevents this scenario. When the speed of second sound has reached its minimum in the upstream direction, it becomes extremely sensitive to further 
changes of the superflow. A tiny change in $v$ dramatically increases $u_2$ until it reaches the speed of first sound. This is the point where one of the modes
becomes unstable, and the other is damped. The angular plots show that this instability extends from the exact upstream direction to almost all 
backward directions $\theta\in \left[\frac{\pi}{2},\frac{3\pi}{2}\right]$, 
however never reaching the half-space of forward directions. At some point, the instability is gone for the anti-parallel direction and only persists
in some nontrivial angular regime. The rapid increase of the upstream sound speed from almost zero ``through'' the instability continues until values larger 
than one are reached. This is particularly obvious in the left panel of Fig.\ \ref{fig:u1u2}, where the two sound speeds have the form of two crossing curves, 
only that the crossing region is replaced by the instability. 

Qualitatively, the picture of Fig.\ \ref{fig:polar1} is valid for all temperatures except for very small temperatures. The corresponding series of polar plots for 
such a small temperature, $T=0.009\,T_c$, is shown in Fig.\ \ref{fig:polar2}. This plot also 
serves as a check for the numerics, because the low-temperature results can be compared to the analytical $T\to 0$ results from Eqs.\ (\ref{u12}): 
the results show that the speed of first sound at $T=0.009\,T_c$ is indistinguishable from the $T\to 0$ result, while the speed of second sound is 
much more sensitive to small changes in temperature, which is already clear from the results without superflow, see Fig.\ \ref{fig:v0}. 
I have checked that upon decreasing the temperature even further, the result for $u_2$ indeed approaches the 
zero-temperature result. The first row of the figure is an extension of the results of Ref.\ \cite{2013PhRvD..87f5001A} (see Fig.\ 1 in that work) to nonzero 
temperatures. In Ref.\ \cite{2013PhRvD..87f5001A}, a series expansion in $T$ was performed. In this series, higher powers in $T$ appear together with higher 
powers in $1/(1-3v^2)$, and thus the expansion breaks down close to the critical velocity $v\to \frac{1}{\sqrt{3}}$. Therefore, for large $v$, only the limit $T\to 0$ 
had been discussed. 

\begin{figure}[t] 
\begin{center}
\hbox{\includegraphics[width=0.265\textwidth]{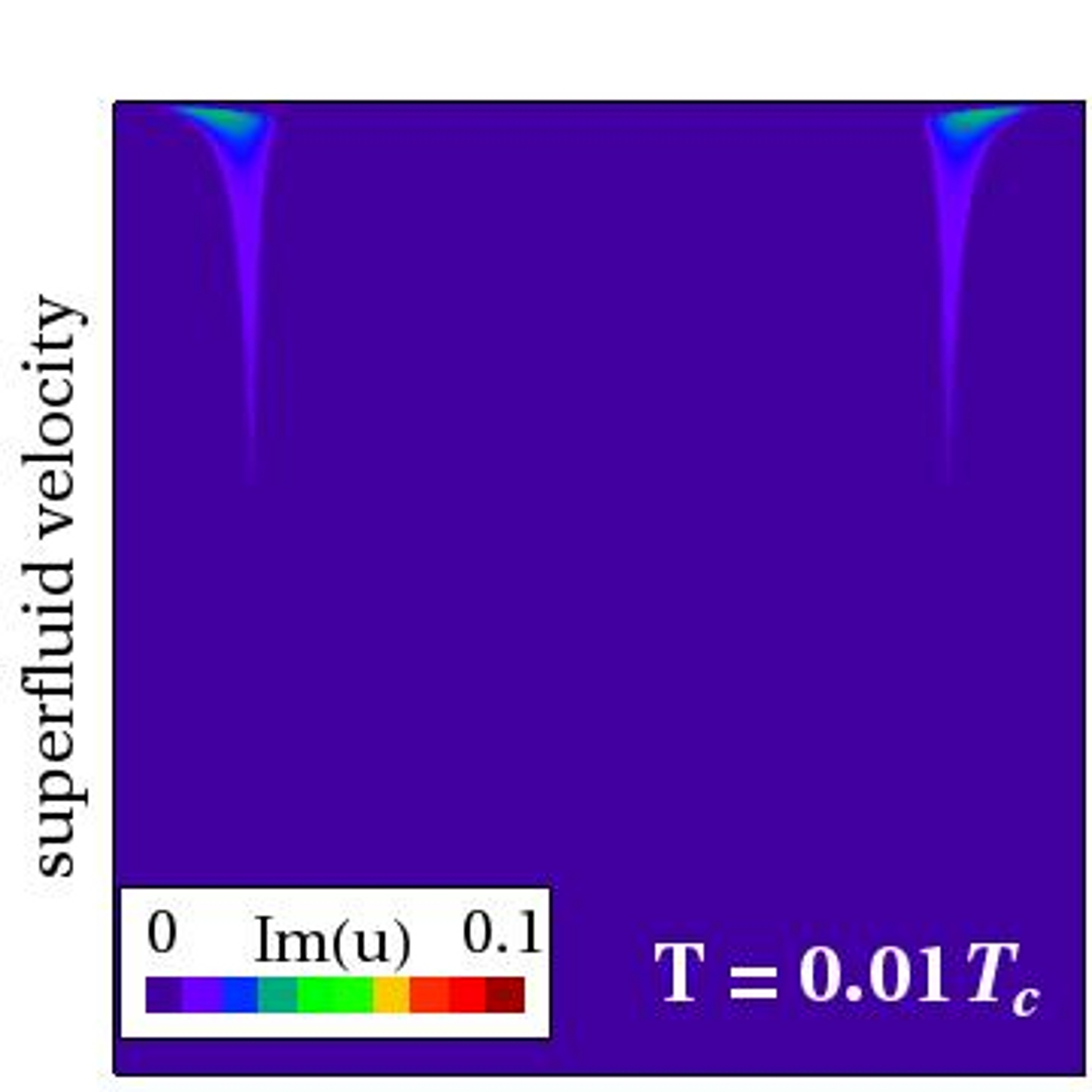}\includegraphics[width=0.24\textwidth]{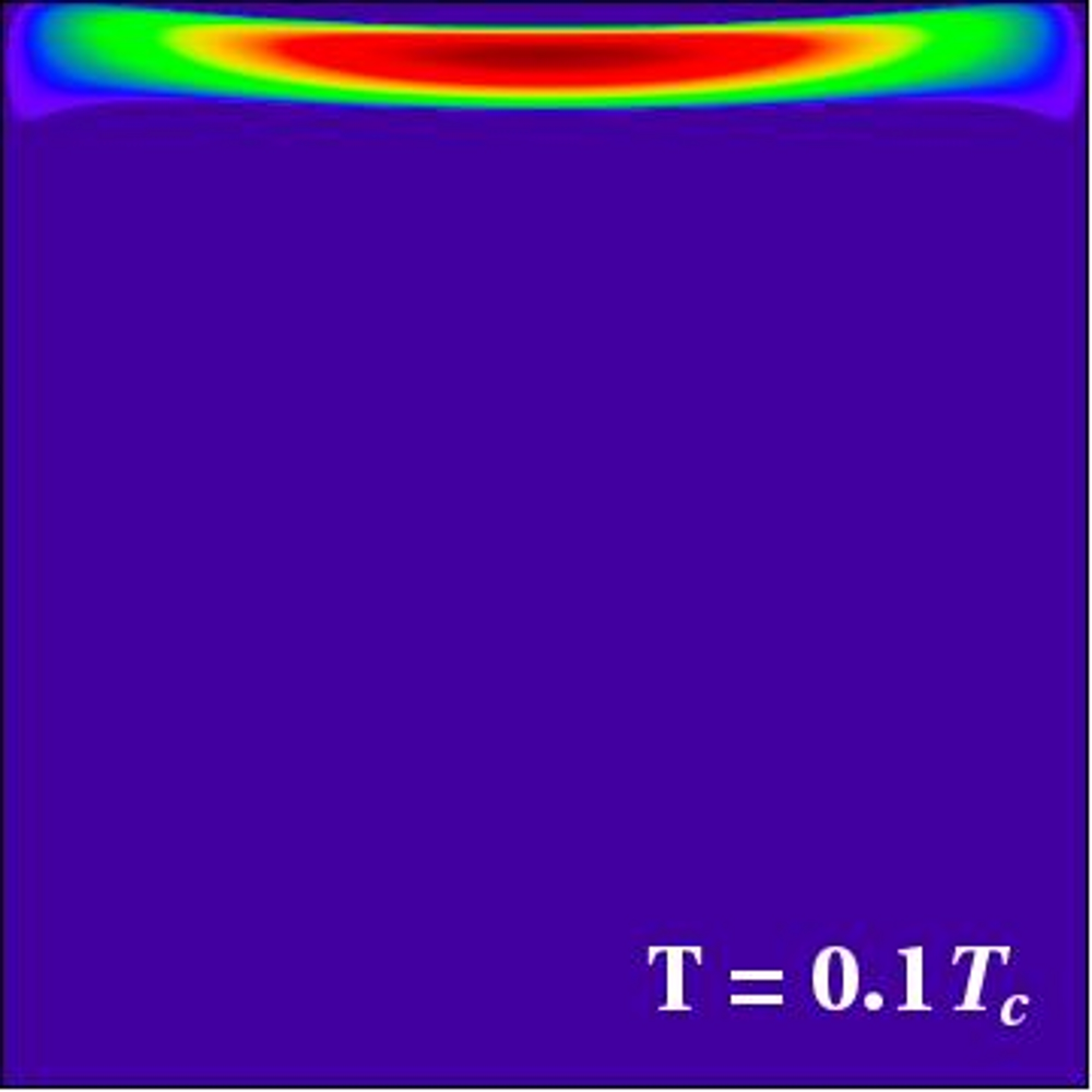}\includegraphics[width=0.24\textwidth]{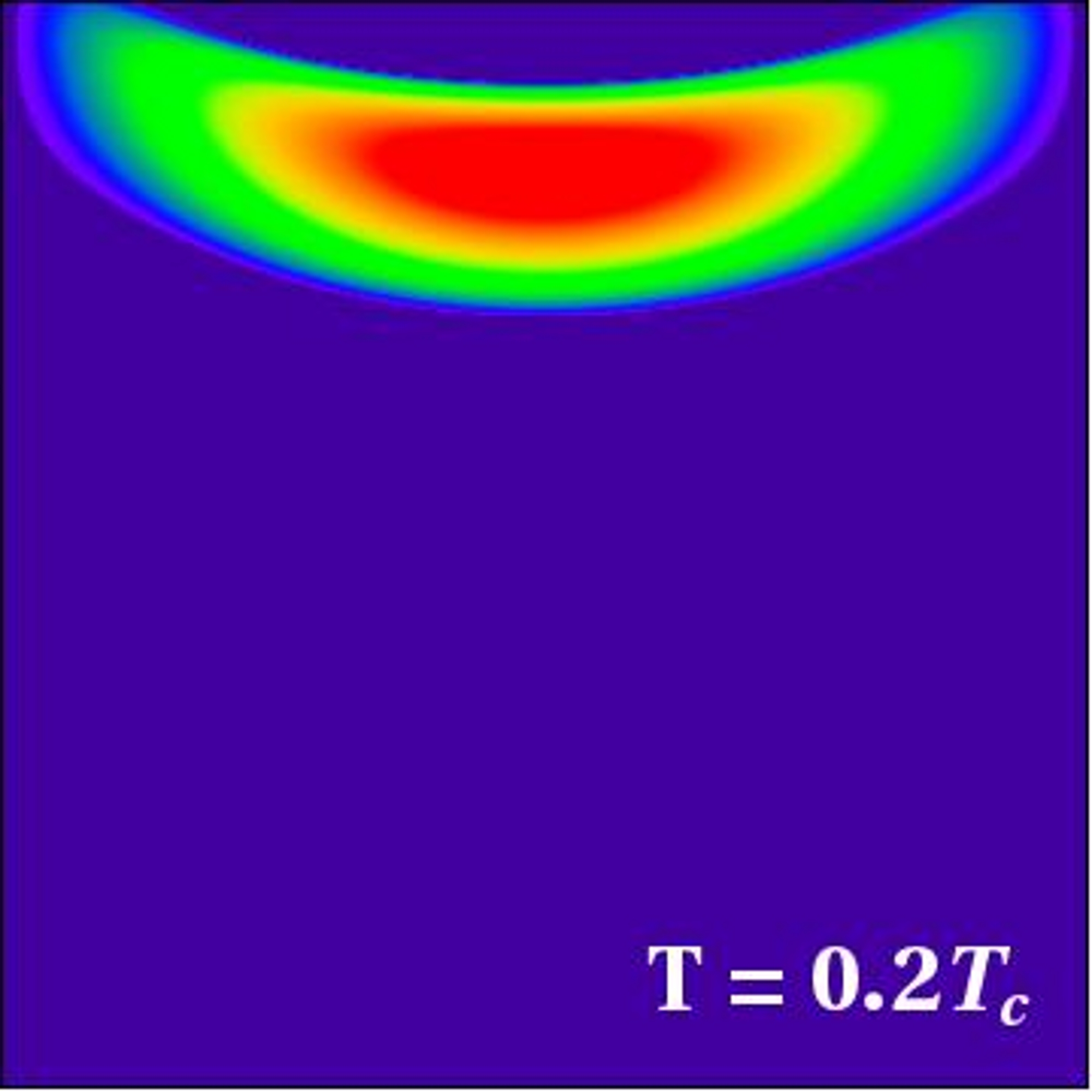}\includegraphics[width=0.24\textwidth]{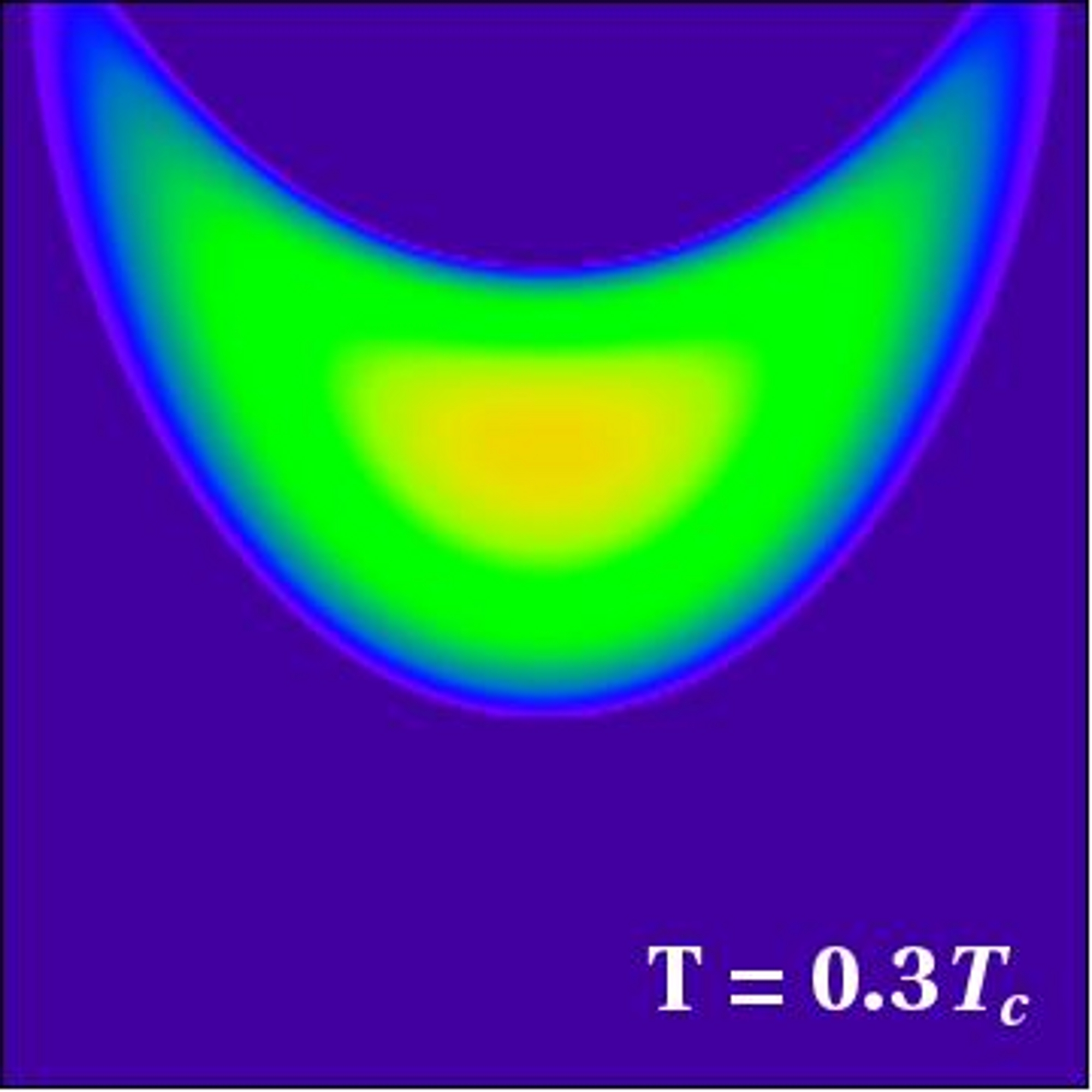}}\hbox{\includegraphics[width=0.265\textwidth]{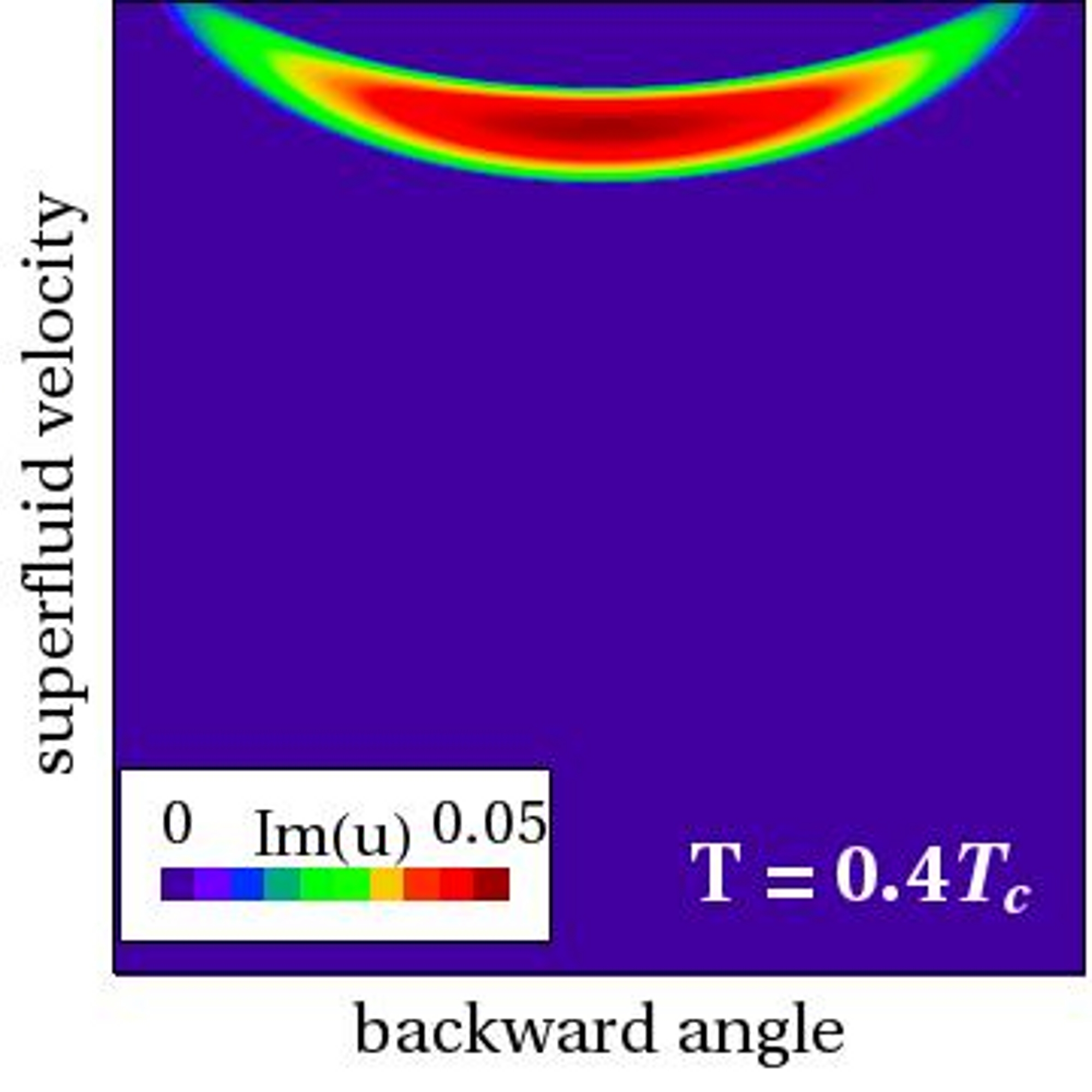}\includegraphics[width=0.24\textwidth]{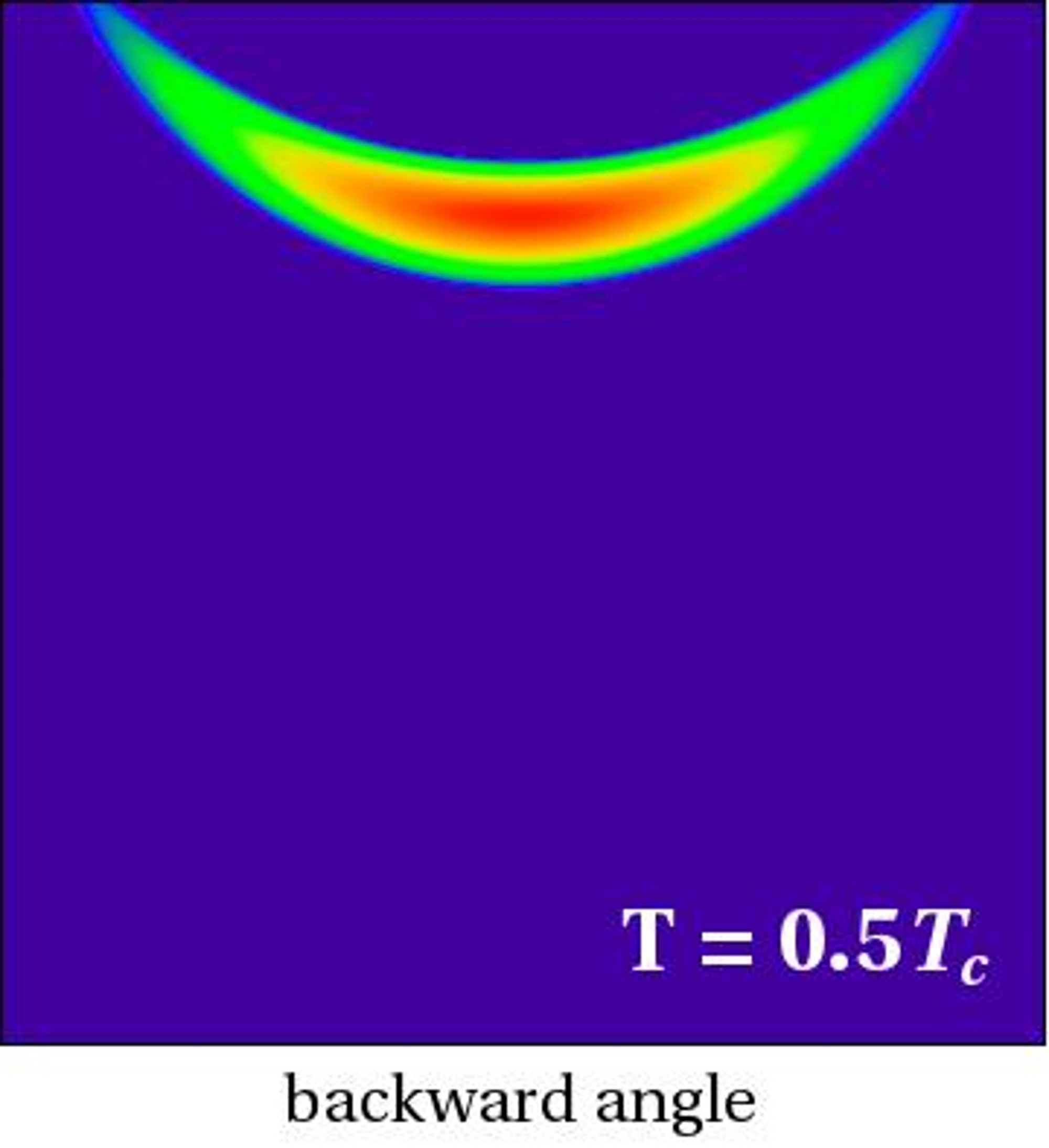}\includegraphics[width=0.24\textwidth]{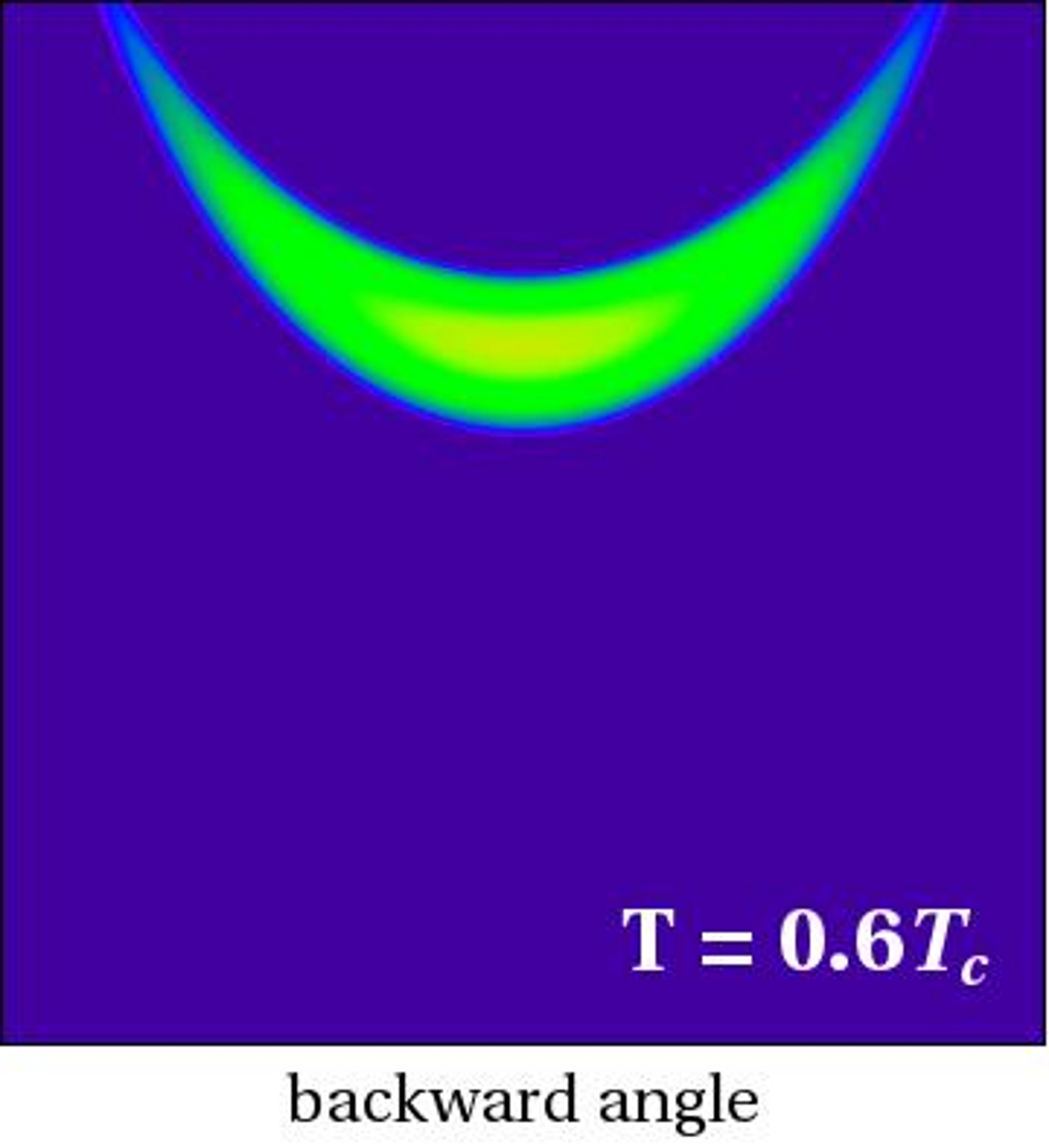}\includegraphics[width=0.24\textwidth]{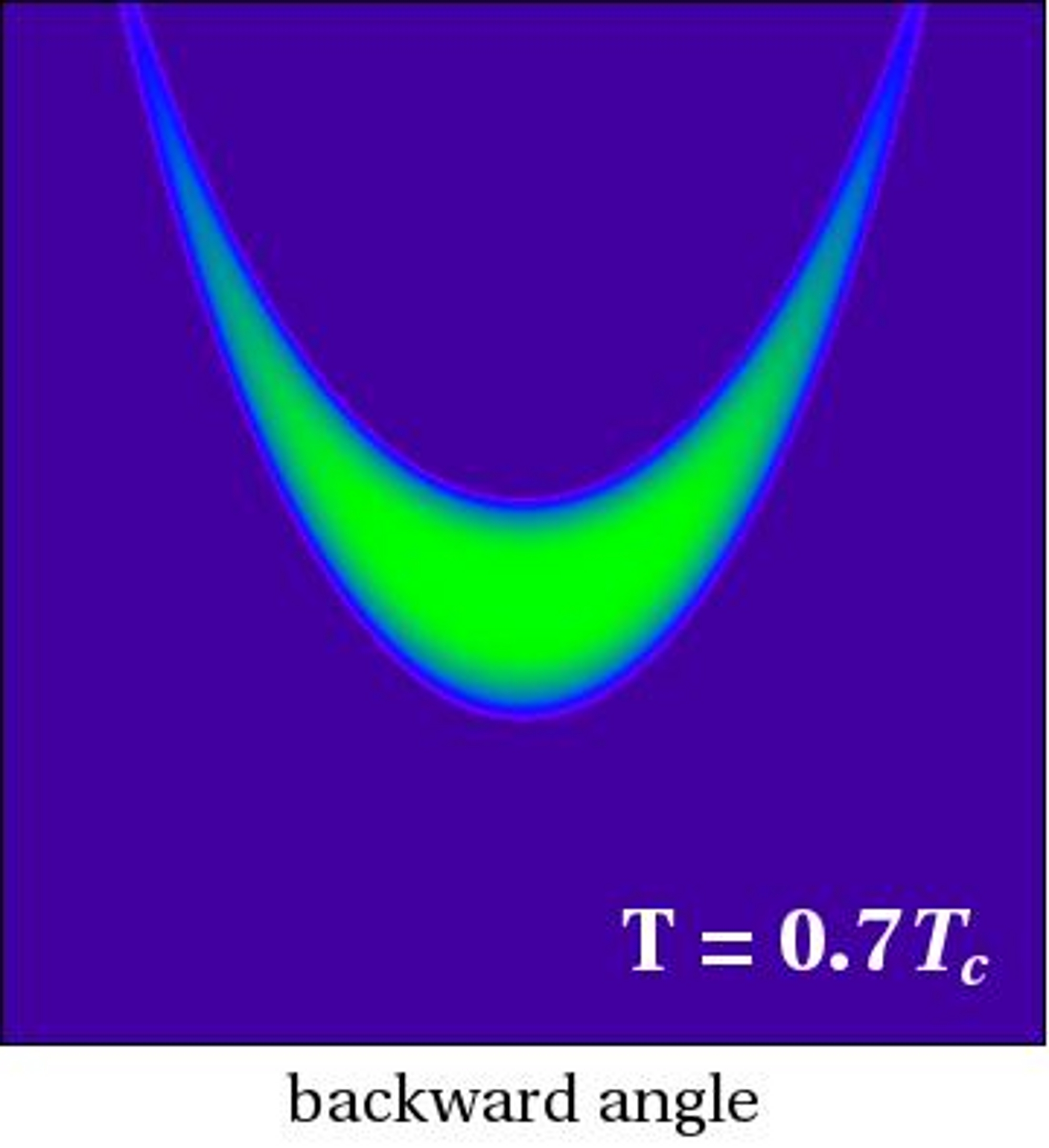}}
\caption{Imaginary part of the sound speeds $u$ as a function of the superfluid velocity $v$ and the angle $\theta\in [\frac{\pi}{2},\frac{3\pi}{2}]$ 
in the backward directions 
with respect to the superflow, for various temperatures. The scale for the superfluid velocity starts at $v=0.995\,v_c(T)$ (upper row), $v=0.985\,v_c(T)$ (lower row), 
and terminates at the point above which $u>1$. Note that the color scale is different for first and second rows. The larger the imaginary part of the sound
mode (for a given momentum of the wave), the faster the unstable mode grows. In this sense, the most severe instability in these plots occurs at $T= 0.1 T_c$. }
\label{fig:angles}
\end{center}
\end{figure}

The situation appears to be similar to larger temperatures in that the speed of second sound approaches zero for the upstream direction
just before the onset of the instability. 
However, the difference to the scenario of Fig.\ \ref{fig:polar1} is that at the point where $u_2(\theta=\pi)\to 0$, there are certain angles where the 
speeds of first and second sound are identical (compare the upper right panels in Figs.\ \ref{fig:polar1} and \ref{fig:polar2} to see the difference). It is 
at these points where the instabilities now set in, while modes along the anti-parallel direction remain stable. 
%This behavior is already indicated by the 
%zero-temperature limit of Sec.\ \ref{sec:zeroT}: in this limit, there is no instability before Landau's critical velocity is reached; but, at Landau's critical
%velocity, the speed of second sound is about to become complex in the directions {\it perpendicular} to the superflow. This suggests that, by decreasing the temperature, 
%the angle where the instability occurs approaches $\theta=\frac{\pi}{2}$. 
The low-temperature instability occurs for superfluid velocities extremely close to $v_c(T)$, pushing the requirements for the accuracy of the numerical 
calculation. I cannot completely exclude that numerical uncertainties affect the results quantitatively in this very-close-to-critical regime. However, the only non-trivial 
numerical operations to be done here are solving the algebraic self-consistency equations and performing (numerous) three-momentum integrals, making the 
evaluation tedious, but numerically very stable. There is no artifact from potentially negative values of the Goldstone dispersion: 
the Goldstone mode in the close-to-critical regime has a very flat low-energy dispersion; but I have checked explicitly 
that the dispersion is still positive for all momenta, as it should be -- by definition -- for superfluid velocities below Landau's critical velocity.

Fig.\ \ref{fig:angles} shows the magnitude of the imaginary part of the sound speeds. While at very low temperatures the instability sets in 
for non-trivial angles, as just pointed out, there is a temperature regime where the instability occurs almost simultaneously for all angles, 
before, for $T\gtrsim 0.2 T_c$, the instability sets in first in the exact upstream direction. Except for very low temperatures, the 
magnitude of the instability (in terms of the magnitude of the imaginary part) decreases monotonically with temperature (note the two different
color scales in Fig.\ \ref{fig:angles}, one for the upper panels, one for the lower ones). Remember that $\gamma = k\,{\rm Im(u)}$, where $k$ is the wave number of the 
sound mode, is the inverse time scale for the exponential growth of the unstable mode. 
The other trend that can be seen is the decreasing superfluid velocity (relative to $v_c(T)$) at which the instability sets in 
(note that the upper and lower panels in Fig.\ \ref{fig:angles} have different offset values
for the vertical scale, as explained in the caption). Again, the low-temperature regime is an exception from this trend. Finally, one can see that $u>1$,
the onset of which is the upper boundary for each of the panels, occurs when the two-stream instability is just about to disappear.

\subsection{Dependence on boson mass and coupling constant}
\label{sec:dependence}

\begin{figure}[t] 
\begin{center}
\hbox{\includegraphics[width=0.48\textwidth]{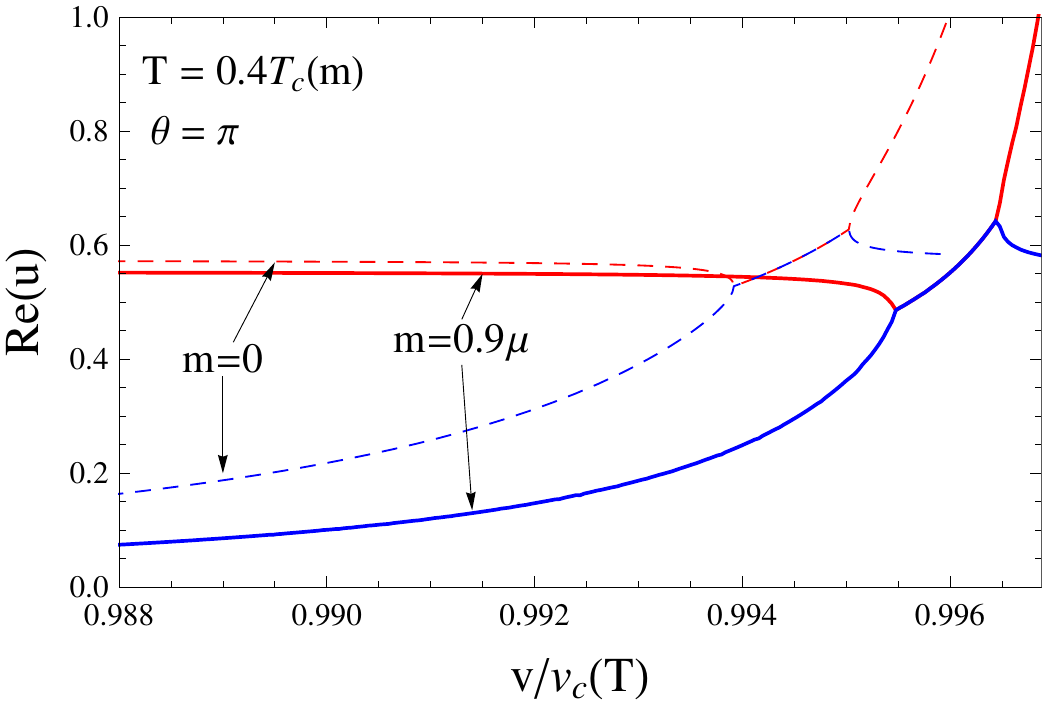}\hspace{0.2cm}\includegraphics[width=0.5\textwidth]{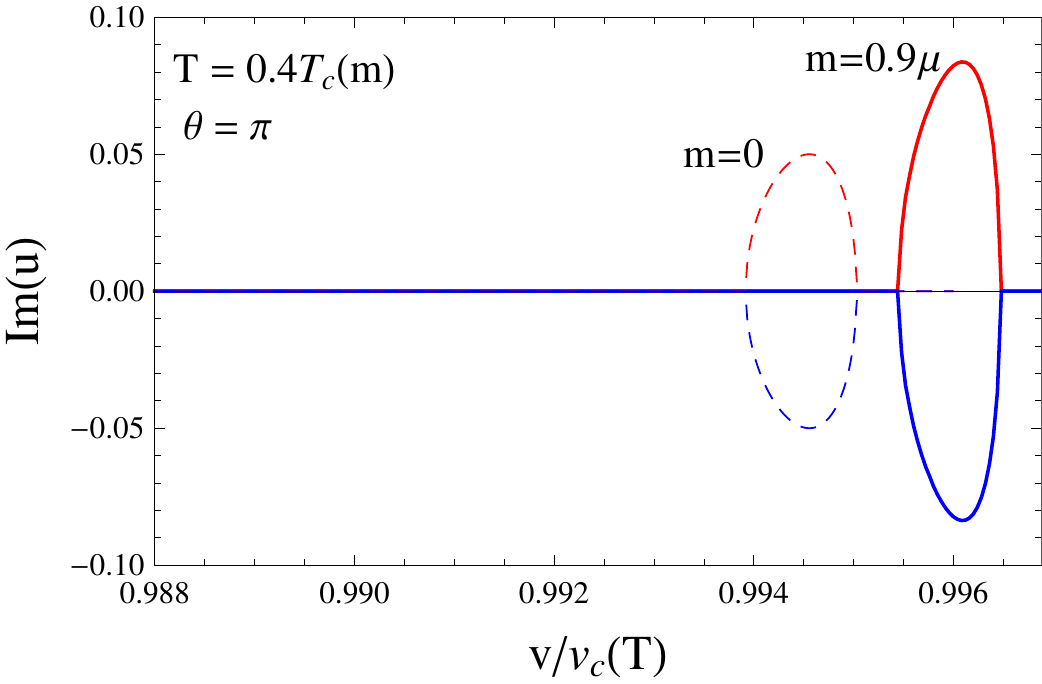}}
\caption{Real (left) and imaginary (right) parts of the sound velocities anti-parallel to the superflow 
for two different boson masses $m=0$ (thin dashed lines) and $m=0.9\mu$ (thick solid lines), at the same {\it relative} temperature $T=0.4\,T_c(m)$
and for a coupling constant $\lambda=0.05$. The ultra-relativistic case suffers the instability already for smaller velocities relative to Landau's
critical velocity $v_c(T)$, but shows a somewhat milder instability (=smaller magnitude of the imaginary part). }
\label{fig:m0m09}
\end{center}
\end{figure}

\begin{figure}[t] 
\begin{center}
\hbox{\includegraphics[width=0.48\textwidth]{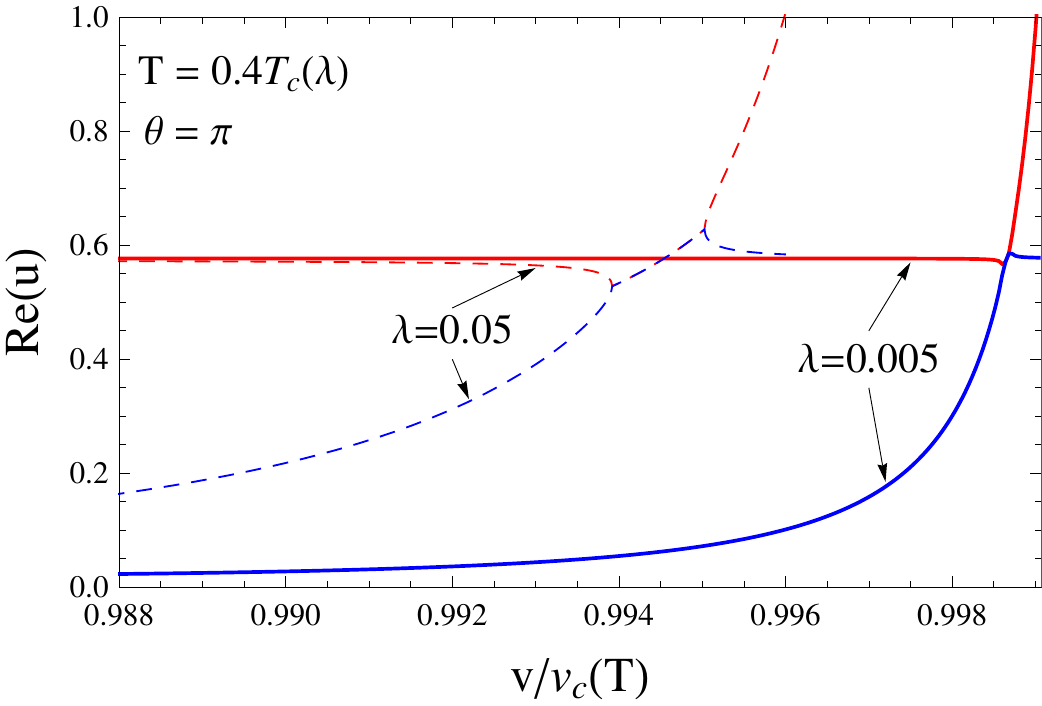}\hspace{0.2cm}\includegraphics[width=0.5\textwidth]{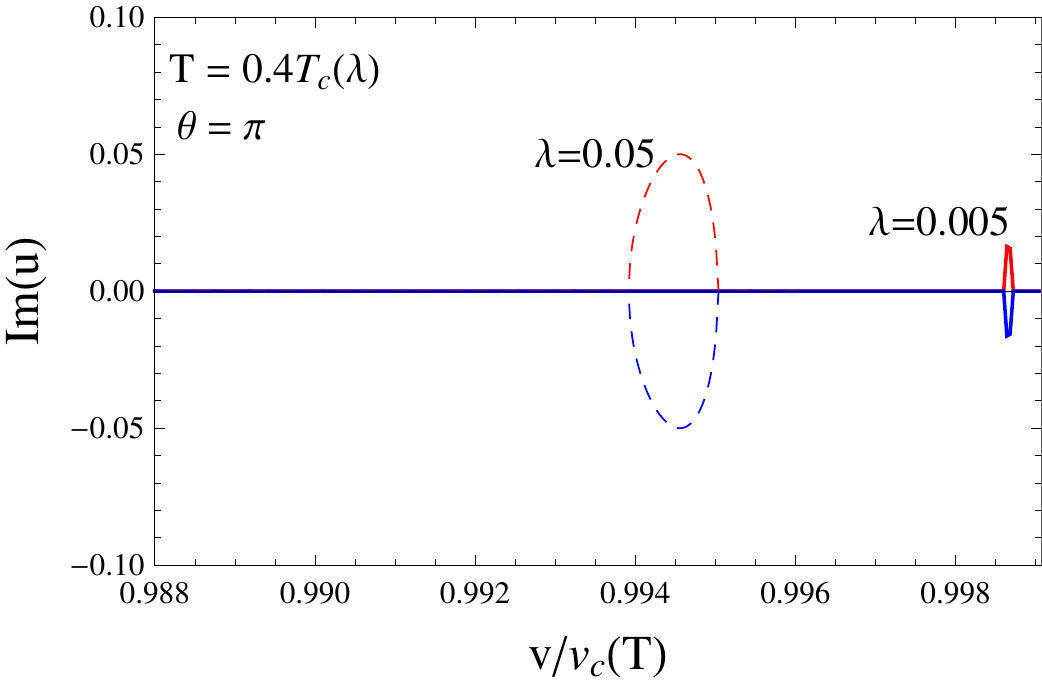}}
\caption{Real (left) and imaginary (right) parts of the sound velocities anti-parallel to the superflow 
for two different coupling constants, $\lambda=0.05$ (thin dashed lines) and $\lambda=0.005$ (thick solid lines), at the same {\it relative} temperature 
$T=0.4\,T_c(\lambda)$, in the ultra-relativistic limit $m=0$. }
\label{fig:lambda}
\end{center}
\end{figure}

The results of the previous sections were all obtained in the ultrarelativistic limit $m=0$. Within the present approach, the instabilities
can be studied for all values of the boson mass, i.e., one can continuously extrapolate from the ultrarelativistic to the non-relativistic limit. 
In the non-relativistic limit, $m$ is very close to, but still below, the chemical potential $\mu$ (only for $m<\mu$ there is condensation). Here I do not  
attempt to present a study of the whole parameter space, I will rather focus on one nonzero value for the boson mass. 
In Fig.\ \ref{fig:m0m09}, real and imaginary parts of the sound velocities in the upstream direction for $m=0.9\mu$ are shown in comparison
to the ultra-relativistic result from Fig.\ \ref{fig:u1u2}. The modes for the two different masses are plotted at the same {\it relative} temperature [with respect to 
the critical temperature in the absence of a superflow $T_c(m)$] versus the {\it relative} velocity
[with respect to Landau's critical velocity $v_c(T)$]. The critical temperatures are $T_c(m)=7.71\mu, 3.32\mu$, and the two 
critical velocities $v_c(0.4T_c(m)) = 0.527,0.228$ for $m=0,0.9\mu$, respectively. 
Even for the larger mass, the sound speeds are still sizable fractions
of the speed of light, i.e., in this sense $m=0.9\mu$ is still far from the non-relativistic limit. It was shown in Ref.\ \cite{Alford:2013koa}, however, that 
already for $m=0.6\mu$ the sound modes show qualitative features identical to a pure non-relativistic calculation. The results of  Fig.\ \ref{fig:m0m09}
suggest that the non-relativistic two-stream instability sets in later, i.e., for larger (relative) superfluid velocities, than the ultra-relativistic two-stream 
instability. 
The magnitude of the imaginary part is larger in the case $m=0.9\mu$, i.e., the unstable mode grows faster in that case.

A similar comparison can be made for the coupling strength. In Fig.\ \ref{fig:lambda}, the real and imaginary parts of  the sound modes are shown for 
$\lambda=0.005$ and compared to the results for the larger coupling $\lambda=0.05$ from the previous sections. One can see that the stronger the 
coupling the more severe the instability: for the smaller coupling, the instability sets in later, i.e., for larger superflows, and the unstable modes 
grow slower. This suggests that a larger microscopic coupling also leads to a larger cross-coupling of the two fluids. 

\subsection{Phase diagram}
\label{sec:phasediagram}

\begin{figure}[t] 
\begin{center}
\hbox{\includegraphics[width=0.48\textwidth]{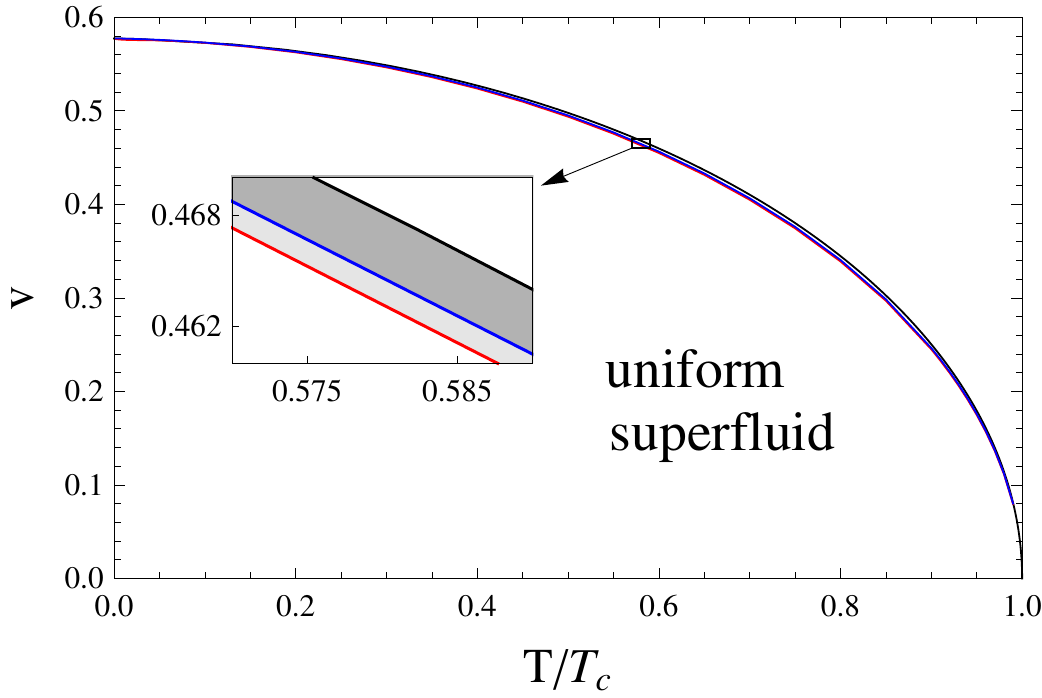}\includegraphics[width=0.5\textwidth]{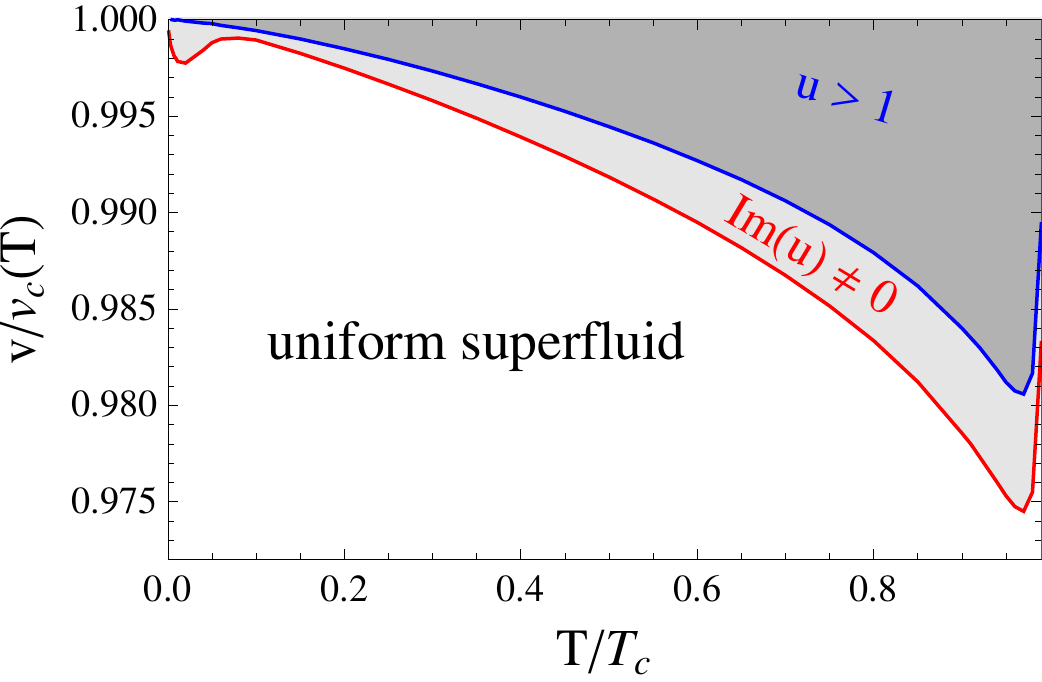}}
\caption{Phase diagram in the plane of superfluid velocity $v$ and temperature $T$ for $m=0$ and $\lambda=0.05$. The left panel illustrates the smallness of the region
where the two-stream instability occurs. The right panel zooms into that region and shows the regime of the two-stream instability where there exists 
a mode with ${\rm Im}(u)\neq 0$ and the unphysical regime of a sound velocity larger than the speed of light, $u>1$. 
Landau's critical velocity, where the Goldstone dispersion becomes negative, is denoted by $v_c(T)$, and $T_c$ is the critical temperature in the absence of a superflow.}
\label{fig:phasediag}
\end{center}
\end{figure}

I will now go back to the parameters used before, $m=0$, $\lambda=0.05$, and discuss the phase diagram in the plane of (uniform) superfluid 
velocity and temperature. In Ref.\ \cite{Alford:2013koa}, such a phase diagram was discussed by computing Landau's critical velocity according to the 
condition (\ref{e0}). The improvement of this phase diagram, by computing the onset of the two-stream instability for all temperatures, is shown in 
Fig.\ \ref{fig:phasediag}. The left panel illustrates that the region covered by the instability is, for the chosen value of the coupling constant, very small. 
For all temperatures, 
the uniform superfluid is stable for all superfluid velocities smaller than about 98\% of Landau's critical velocity. In order to study the unstable region in more 
detail, the right panel zooms into this region. Note that the curve  $v_c(T)$ of the left panel is identical to the upper horizontal border of the right panel. 
The region where the two-stream instability operates [labelled by ``${\rm Im}(u)\neq 0$''] is defined such that for any point in this region 
there exists at least one spatial direction in which one sound mode is unstable. At the lower critical line, which has to be crossed to enter this region, 
this is the direction anti-parallel to the superflow for most temperatures, except for very small temperatures, where 
the instability occurs for some non-trivial angle, as discussed above. This qualitative difference manifests itself in the phase transition line, which shows a 
non-monotonic behavior for $T\lesssim 0.1\,T_c$. Also close to $T_c$, the phase transition line is non-monotonic. However, in this regime, artifacts of the 
Hartree approximation may play an important role, and thus this behavior should be taken with some care.  

%In the region labelled by ``$u>1$'' there exists at least one direction in which there is a sound speed that exceeds the speed of light. 
%While the two-stream-unstable region yields a physical indication of what the system wants to do, it is unclear whether there is a physical meaning 
%behind $u>1$, or whether this behavior is an artifact of the approximation scheme used here. As mentioned above, I 
%have checked that the unphysical behavior is only seen in the sound modes: all thermodynamic quantities are well behaved. One might ask what the system would do 
%if we placed it in the ``acausal'' region. It is, however, not obvious how to bring the system in this region: entering from below, say by increasing the superflow
%at a fixed temperature, will run into the two-stream instability, and the system can no longer be described in the present approach. Entering from above,
%say by cooling at a fixed superflow is not possible either: within the given uniform, dissipationless ansatz, the points in the phase diagram above Landau's critical 
%velocity are meaningless, there is no stable state. It will thus be interesting to see in future studies whether the problem of sound speeds larger than the 
%speed of light still occurs in the presence of dissipation and/or inhomogenhous condensates. 

\section{Conclusions}
\label{sec:conclusions}

I have computed real and imaginary parts of the two sound velocities in a relativistic, dissipationless superfluid, using a microscopic field-theoretical model.
This model contains a complex scalar 
field with quartic self-interaction. Most of the results have been computed in the ultra-relativistic, weak-coupling limit.
The microscopic calculation involves solving a self-consistent 
equation for the Bose-Einstein condensate that spontaneously breaks the $U(1)$ symmetry of the underlying Lagrangian. This self-consistent
equation, together with a Dyson-Schwinger equation for the boson propagator, is obtained from the two-particle-irreducible effective action. 
The sound velocities, derived from the linearized hydrodynamic equations, 
are entirely determined by thermodynamic equilibrium quantities, computed from the field theory. The thermodynamic ensemble
is given by temperature $T$, the chemical potential $\mu$ associated to the $U(1)$ charge, and a uniform superfluid velocity ${\bf v}$.
The superfluid velocity is the velocity of the superfluid measured in the rest frame of the normal fluid, in which the field-theoretical calculation is
performed. This externally given superflow is the crucial ingredient because it gives rise to a two-stream system whose two constituents interact.   

The main result is the occurrence of nonzero imaginary parts of the sound velocities. These imaginary parts have opposite signs for the two sound modes, indicating
one exponentially growing mode whose existence can be interpreted as the two-stream instability \cite{Samuelsson:2009up}.  This dynamical 
instability occurs for all nonzero 
temperatures and for superfluid velocities very slightly below Landau's critical velocity, which is defined by the onset of negative quasiparticle 
energies. As a consequence, there is a small band in the $T$-$v$ phase diagram where the two-stream instability operates, its width varying from zero at zero temperature
up to about 2.5\% of Landau's critical velocity at temperatures near $T_c$, the critical temperature in the absence of a superflow. 
If the superfluid velocity is further increased within the dynamically unstable region, one of the sound speeds becomes larger than the speed of light. 
Possibly, this indicates a problem with the underlying formalism. Therefore, it would be interesting to repeat the calculation in a more complete treatment, 
for instance by going beyond the Hartree approximation or by implementing the Goldstone theorem differently, or by using a different microscopic model. 

I have discussed the full angular dependence of the two-stream instability, showing that it typically sets in first (= for the lowest superflow) 
for sound modes propagating in the exact upstream direction. In this direction, the speed of second sound approaches zero, i.e., ``wants'' to move together 
with the normal fluid component,
before the instability sets in. In a way, the instability appears to prevent the upstream sound wave to become a downstream sound wave, in which case two observers
in the two rest frames of normal fluid and superfluid would see it propagate in opposite directions. At small temperatures, a somewhat different behavior is observed. 
For $T\lesssim 0.1\,T_c$, the instability only operates at some non-trivial backward angle, and no unstable mode is seen in the exact upstream direction. 
In this particular backward angle, first and second sound have the same velocity, which suggests that there is some efficient energy transfer from first to second sound
that triggers the instability.

The present model allows us to extrapolate continuously from the ultra-relativistic to the non-relativistic limit by varying the boson mass. For a larger mass the 
unstable region in the phase diagram appears to become smaller (relatively speaking, i.e., compared to the stable region), but the instability seems to be 
slightly more severe since the time scale for the growth of the unstable mode becomes somewhat shorter. One can also vary the coupling constant of the 
underlying microscopic field theory. The instability becomes weaker for smaller coupling:
both the size of the unstable region of the phase diagram and the strength in terms of the magnitude of the imaginary part of the sound speed 
decrease for smaller coupling. It would thus be very 
interesting to repeat the same calculation for larger coupling strengths and see whether the phase diagram then is covered by a more sizable unstable region. 
In the given model, this would require to resolve the difficulties related to the renormalization of the approach in the presence of a nonzero
superflow \cite{Alford:2013koa}. Another way of testing the strong-coupling behavior would be the gauge/gravity duality. A phase diagram including Landau's critical 
velocity, but without calculations of the sound modes that could reveal the two-stream instability, has recently been obtained in such a holographic 
approach \cite{Amado:2013aea}.

The results of this paper cannot predict the real-time evolution of the superfluid two-stream instability and, due to the restriction to weak 
coupling, are most likely not directly applicable to realistic superfluids in compact stars or in the laboratory. Extensions along both lines are therefore interesting
projects for the future. For a recent real-time  simulation in a general hydrodynamic multi-fluid setup, not referring to a microscopic model, 
see Ref.\ \cite{Hawke:2013haa}.  In particular in the astrophysical context, where superfluids in ultra-dense quark or nuclear matter require 
a relativistic treatment, the present calculation and its extensions may become relevant. One can ask how a two-stream instability 
manifests itself in a compact star, whether and how it is damped, and whether it may help to understand pulsar glitches \cite{2004MNRAS.354..101A}. 

\begin{acknowledgments}
I am grateful to G.\ Comer for drawing my attention to the superfluid two-stream instability. I thank M.\ Alford, S.K.\ Mallavarapu, S.\ Stetina for the collaboration on 
our previous project upon which the present results are built and  A.\ Rebhan for useful comments and discussions.
This work has been supported by the Austrian science foundation FWF under project no.~P23536-N16 and by the NewCompStar network, COST Action MP1304.
\end{acknowledgments}

\appendix

\section{Polynomial for sound velocities}
\label{appA}

The coefficients appearing in Eq.\ (\ref{matrix}) are given by
\begin{subequations}
\bea
a_1 &\equiv& \frac{w}{s}\frac{\partial n}{\partial T} \, , \qquad a_2 \equiv -n_n \,  , \qquad 
a_3\equiv \frac{n_s}{\sigma}-\frac{w}{s}\frac{\partial (n_s/\sigma)}{\partial T} +\frac{n_n}{s}\frac{\partial n}{\partial T}-\frac{\partial n}{\partial \mu}
-2\mu\frac{\partial n}{\partial (\nabla\psi)^2} \, , \non[2ex]
a_4&\equiv&-\left[\frac{n_n}{s}\frac{\partial (n_s/\sigma)}{\partial T}-\frac{\partial (n_s/\sigma)}{\partial \mu}-2\mu\frac{\partial (n_s/\sigma)}
{\partial (\nabla\psi)^2}\right] \, , \allowdisplaybreaks\\[2ex]
b_1 &\equiv& \frac{w}{s}\frac{\partial s}{\partial T} \, , \qquad b_2\equiv-s \, , \qquad b_3 \equiv \frac{n_n}{s}\frac{\partial s}{\partial T}-\frac{\partial s}{\partial \mu}
-2\mu\frac{\partial s}{\partial (\nabla\psi)^2} \, , \allowdisplaybreaks\\[2ex]
A_1 &\equiv& \mu\frac{\partial n}{\partial\mu}+T\frac{\partial n}{\partial T} \, , \qquad A_2 \equiv -n \, ,\qquad 
A_3\equiv \frac{n_s}{\sigma}-\mu\frac{\partial(n_s/\sigma)}{\partial\mu}-T\frac{\partial(n_s/\sigma)}{\partial T}+\frac{n_n}{s}\frac{\partial n}{\partial T}-
\frac{\partial n}{\partial \mu} \, , \non[2ex]
A_4&\equiv&-\left[\frac{n_n}{s}\frac{\partial (n_s/\sigma)}{\partial T}-\frac{\partial (n_s/\sigma)}{\partial \mu}\right] \, ,  \allowdisplaybreaks\\[2ex]
B_1 &\equiv& \mu\frac{\partial s}{\partial\mu}+T\frac{\partial s}{\partial T} \, , \qquad B_2\equiv-s \, , \qquad B_3 \equiv \frac{n_n}{s}\frac{\partial s}{\partial T}
-\frac{\partial s}{\partial \mu} \, .
\eea
\end{subequations}
The coefficients of the polynomial (\ref{poly}) are 
\begin{subequations}
\bea
Q^{(4)} &=& \frac{\mu w}{s}\left(\frac{\partial n}{\partial T}\frac{\partial s}{\partial \mu}-\frac{\partial s}{\partial T}\frac{\partial n}{\partial \mu}\right) 
\allowdisplaybreaks\\[2ex]
Q^{(3)} &=& 2\mu\frac{\partial n}{\partial(\nabla\psi)^2} d_1(s)
-2\mu\frac{\partial s}{\partial(\nabla\psi)^2}d_1(n)
+\frac{\mu w}{s}\left(\frac{\partial (n_s/\sigma)}{\partial T}\frac{\partial s}{\partial \mu}-\frac{\partial (n_s/\sigma)}{\partial \mu}
\frac{\partial s}{\partial T}\right) 
-\frac{\mu n_s}{\sigma} d_2(s) \, , \allowdisplaybreaks\\[2ex]
Q^{(2)}_{1}&=&-2n_n\mu\frac{\partial s}{\partial \mu} +s\mu\frac{\partial n}{\partial \mu}+\left(\frac{nw}{s}-n_nT\right)\frac{\partial s}{\partial T} \, , 
\allowdisplaybreaks\\[2ex]
Q^{(2)}_{2}&=&2\mu\frac{\partial n}{\partial(\nabla\psi)^2} d_2(s)
+2\mu\frac{\partial (n_s/\sigma)}{\partial(\nabla\psi)^2} d_1(s) 
+2\mu\frac{\partial s}{\partial(\nabla\psi)^2}\left(\frac{n_s}{\sigma}-d_1(n_s/\sigma)-d_2(n)\right) \, , \allowdisplaybreaks\\[2ex]
Q^{(1)}_{1}&=& 2\mu n \frac{\partial s}{\partial(\nabla\psi)^2}- 2\mu s \frac{\partial n}{\partial(\nabla\psi)^2} +\frac{\mu n_s}{\sigma} d_2(s)
+\mu s d_2(n_s/\sigma) \, , \allowdisplaybreaks\\[2ex]
Q^{(1)}_{2} &=& 2\mu\frac{\partial (n_s/\sigma)}{\partial(\nabla\psi)^2} d_2(s)
-2\mu\frac{\partial s}{\partial(\nabla\psi)^2} d_2(n_s/\sigma) \, , \allowdisplaybreaks\\[2ex]
Q^{(0)}_{1} &=& -\frac{\mu n_s}{\sigma}s \, , \allowdisplaybreaks\\[2ex] 
Q^{(0)}_{2} &=& -2\mu s\frac{\partial (n_s/\sigma)}{\partial(\nabla\psi)^2} \, .
\eea 
\end{subequations}
where I have used the abbreviations 
\bea
d_1(x)\equiv \mu\frac{\partial x}{\partial \mu}+T\frac{\partial x}{\partial T} \, , \qquad 
d_2(x)\equiv \frac{\partial x}{\partial \mu}-\frac{n_n}{s}\frac{\partial x}{\partial T} \, .
\eea

\bibliography{refs}

\end{document}